\documentclass[12pt,reqno]{amsart}
\usepackage{graphicx}
\usepackage{amscd}
\usepackage{amsmath}
\usepackage{epsfig}
\usepackage{amsfonts}
\usepackage{amssymb}

\providecommand{\U}[1]{\protect\rule{.1in}{.1in}}
\providecommand{\U}[1]{\protect\rule{.1in}{.1in}}
\allowdisplaybreaks

\theoremstyle{plain}

\numberwithin{equation}{section}

\input{tcilatex}

\begin{document}
\title[Minimum-Uncertainty Squeezed States]{The Minimum-Uncertainty Squeezed
States \\
for Atoms and Photons in a Cavity}
\author{Sergey I. Kryuchkov}
\address{Mathematical, Computational and Modeling Sciences Center, Arizona
State University, Tempe, AZ 85287--1904, U.S.A.}
\email{sergeykryuchkov@yahoo.com}
\author{Sergei K. Suslov}
\address{School of Mathematical and Statistical Sciences, Arizona State
University, Tempe, AZ 85287--1804, U.S.A.}
\email{sks@asu.edu}
\urladdr{http://hahn.la.asu.edu/\symbol{126}suslov/index.html}
\author{Jos\'{e} M. Vega-Guzm\'{a}n}
\address{Mathematical, Computational and Modeling Sciences Center, Arizona
State University, Tempe, AZ 85287--1904, U.S.A.}
\email{jmvega@asu.edu}
\date{March 25, 2013}

\begin{abstract}
We describe a multi-parameter family of the minimum-uncertainty squeezed
states for the harmonic oscillator in nonrelativistic quantum mechanics.
They are derived by the action of corresponding maximal kinematical
invariance group on the standard ground state solution. We show that the
product of the variances attains the required minimum value $1/4$ only at
the instances that one variance is a minimum and the other is a maximum,
when the squeezing of one of the variances occurs. The generalized coherent
states are explicitly constructed and their Wigner function is studied. The
overlap coefficients between the squeezed, or generalized harmonic, and the
Fock states are explicitly evaluated in terms of hypergeometric functions
and the corresponding photon statistics are discussed. Some applications to
quantum optics, cavity quantum electrodynamics, and superfocusing in
channeling scattering are mentioned. Explicit solutions of the Heisenberg
equations for radiation field operators with squeezing are found.
\end{abstract}

\maketitle

\section{An Introduction}

From the very beginning, nonclassical states of the linear Planck
oscillator, in particular the coherent and squeezed states, have been a
subject of considerable interest in quantum physics (see \cite{Dodonov02}, 
\cite{DodonovManko03}, \cite{GlauberCollect}, \cite{Kennard27}, \cite%
{KlauderSudarshan}, \cite{SchroedingerOscillator}, \cite{Schroedinger} and
the references therein). They occur naturally on an atomic scale \cite%
{Buchleitneretal02}, \cite{Johanningetal09} and, possibly, can be observed
among vibrational modes of crystals and molecules \cite{Demkov09}, \cite%
{DemkovMeyer04}, \cite{Dunnetal95}, \cite{Fey:Hib}. A single monochromatic
mode of light also represents a harmonic oscillator system for which
nonclassical states can be generated very efficiently by using the
interaction of laser light with nonlinear optical media \cite%
{Breit:Schill:Mlyn97}, \cite{LeonardPaul95}, \cite{LvovskyBabichev02}, \cite%
{LvovskyHansenetal01}, \cite{LvovskyMlynek02}, \cite{LvRay09}, \cite%
{RiesetalLv03}, \cite{SchillerBreitenbachetal96}, \cite{Walls83}. Generation
of squeezed light with a single atom has been experimentally demonstrated 
\cite{Ourjoumtsevetal11}. On a macroscopic scale, the squeezed states are
utilized for detection of gravitational waves \cite{Hollenhorst79} below the
so-called vacuum noise level and without violation of the uncertainty
relation \cite{Abadetal11}, \cite{Eberleetal10}, \cite{Pikovskietal12}, \cite%
{Vahlbruch08}.

The past decades progress in generation of pure quantum states of motion of
trapped particles provides not only a clear illustration of basic principles
of quantum mechanics, but it also manifests the ultimate control of particle
motion. These states are of interest from the standpoint of quantum
measurement concepts and facilitate other applications including quantum
computation (see \cite{Bouchouleetal99}, \cite{Cirac12}, \cite%
{CiracBlattZoller94}, \cite{CiracZoller95}, \cite{Gerhardtetal09}, \cite%
{HarocheRaimond06}, \cite{HeinzenWineland90}, \cite{Johanningetal09}, \cite%
{LeibfriedetalWineland03}, \cite{Meekhofetal96}, \cite{Monroeetal95}, \cite%
{Morinagaetal99}, \cite{Paul90}, \cite{Roosetal99}, \cite{Santosetal07} and
the references therein).

It is well known that the harmonic quantum states can be analyzed through
the dynamics of a single, two-level atom which radiatively couples to the
single mode radiation field in the Jaynes--Cummings(--Paul) model \cite%
{CiracBlattetal94}, \cite{Chumakovetal03}, \cite{JaynesCummings63}, \cite%
{LeibfriedetalWineland03}, \cite{Schleich01}, \cite{ShoreKnight93}, \cite%
{VogelFilho95} extensively studied in the cavity QED \cite{DutraQED}, \cite%
{HarocheRaimond06}, \cite{Rempeetal90}, \cite{RempeWaltherKlein87}. Creation
and detection of thermal, Fock, coherent, and squeezed states of motion of a
single $^{9}Be^{+}$ ion confined in a rf Paul trap was reported in \cite%
{Meekhofetal96}, where the state of atomic motion\ had been observed through
the evolution of the atom's internal levels (e.g., collapse and revival)
under the influence of a Jaynes--Cummings interaction realized with the
application of external (classical) fields. The distribution over the Fock
states is deduced from an analysis of Rabi oscillations.

Moreover, Fock, coherent, and squeezed states of motion of a harmonically
bound cold cesium atoms were experimentally observed in a $1D$ optical
lattice \cite{Bouchouleetal99}, \cite{Morinagaetal99}. This method gives a
direct access to the momentum distribution through the square of the modulus
of the wavefunction in velocity space (see also \cite{CiracBlattetal93}, 
\cite{CiracBlattetal94}, \cite{Ciracetal93}, \cite{Cooketal85}, \cite%
{Diedrichetal89}, \cite{HeinzenWineland90}, \cite{Jessenetal92}, \cite%
{Johanningetal09}, \cite{LeibfriedetalWineland03}, \cite{Paul90}, \cite%
{Verkerketal92} and the references therein regarding cold trapped ions and
their nonclassical states; progress in atomic physics and quantum optics
using superconducting circuits is reviewed in \cite{Fu:Mat:Hat:Kur:Zeil}, 
\cite{You:Nori11}).

Recent reports on observations of the dynamical Casimir effect \cite%
{Lahetal11}, \cite{Wilsonetal11} strengthen the interest to the nonclassical
states of generalized harmonic oscillators \cite{Dodonov02}, \cite{Dodonov10}%
, \cite{Dod:Mal:Man75}, \cite{Dod:Man79}, \cite{Dodonov:Man'koFIAN87}, \cite%
{Har:Ben-Ar:Mann11}, \cite{Malkin:Man'ko79}, \cite{Man'koCasimir}, \cite%
{Nationetal12}, \cite{Suslov11} and \cite{Vysotskii2Adamenko12}. The
amplification of quantum fluctuations by modulating parameters of an
oscillator is closely related to the process of particle production in
quantum fields \cite{Dodonov10}, \cite{Jacobson04}, \cite{Man'koCasimir},
and \cite{Nationetal12}. Other dynamical amplification mechanisms include
the Unruh effect \cite{Unruh76} and Hawking radiation \cite{BirrelDavies82}, 
\cite{Hawking74}, \cite{Hawking75}.

The purpose of this paper is to construct the minimum-uncertainty squeezed
states for quantum harmonic oscillators, which are important in these
applications, in the most simple closed form. Our approach reveals the
quantum numbers/integrals of motion of the squeezed states in terms of
solution of certain Ermakov-type system \cite{Lop:Sus:VegaGroup}, \cite%
{LopSusVegaHarm}. The corresponding generalizations of Fock states, which
were originally found in \cite{Marhic78} and recently rediscovered in \cite%
{LopSusVegaHarm}, are discussed in detail. As a result, the probability
amplitudes of these nonclassical states of motion are explicitly evaluated
in terms of hypergeometric functions. Their experimental observations in
cavity QED and quantum optics are briefly reviewed. Moreover, the radiation
field operators of squeezed photons, which can be created from the QED
vacuum, are introduced by second quantization with the aid of hidden
symmetry of harmonic oscillator problem in the Heisenberg picture.

In summary, experimental recognitions of the nonclassical harmonic states of
motion have been achieved through reconstruction of the Wigner function in
optical quantum-state tomography \cite{Breit:Schill:Mlyn97}, \cite{LvRay09},
from a Fourier analysis of Rabi oscillations of a trapped atom \cite%
{Meekhofetal96}, and/or by a direct observation of the square of the modulus
of the wavefunction for a large sample of cold cesium atoms in a $1D$
optical lattice \cite{Bouchouleetal99}, \cite{Morinagaetal99}. Our
theoretical consideration complements all of these advanced experimental
techniques by identifying the state quantum numbers from first principles.
This approach may provide a guidance for engineering more advanced
nonclassical states.

The paper is organized as follows. In sections~2 and 3, we describe the
minimum-uncertainty squeezed states for the linear harmonic oscillator in
coordinate representation. The generalized coherent, or TCS states, are
constructed in section~4. In sections~4 and 5, the Wigner and Moyal
functions of the squeezed states are evaluated directly from the
corresponding wavefunctions and their classical time evolution is verified
with the help of a computer algebra system. The eigenfunction expansions of
the squeezed (or generalized harmonic) states in terms of the standard Fock
ones are derived in section~6 (see also \cite{Dod:Kur:Man80}, \cite%
{Dod:Man:Man94}, \cite{Kimetal89} and the references therein for important
special cases). Some experiments on engineering of \ nonclassical states of
motion are analyzed in section~7. Here, the experimentally observed
probability distributions are derived from our explicit expression for the
probability amplitudes obtained in the previous section. Theoretically
predicted in \cite{Demkov09}, \cite{DemkovMeyer04}, superfocusing in channel
scattering is also discussed. In section~8, we revisit the radiation field
quantization in a perfect cavity, which is important for applications to
quantum optics. Nonstandard solutions of the Heisenberg equations of motion
for the electromagnetic field operators, that naturally describe squeezing
in the Heisenberg picture, are found. The variance of the number operator,
which together with the eigenfunction expansion allows one to compare our
results with experimentally observed squeezed photon statistics \cite%
{Breit:Schill:Mlyn97}, \cite{SchillerBreitenbachetal96}, is evaluated from
first principles in section~9. A brief summary is provided in the last
section. A compact complex parametrization of the Schr\"{o}dinger group can
be found in appendix.

\section{The Minimum-Uncertainty Squeezed States}

The Heisenberg Uncertainty Principle is one of the fundamental laws of
nature and the coherent states that minimize this uncertainty relation are
well known. But, equally important in recent developments,
minimum-uncertainty squeezed states are not so familiar outside a relatively
narrow group of experts. Here, for the benefits of the reader, we construct
these states as explicitly as possible and elaborate on some of their
remarkable features.

The time-dependent Schr\"{o}dinger equation for the simple harmonic
oscillator in one dimension,%
\begin{equation}
2i\psi _{t}+\psi _{xx}-x^{2}\psi =0,  \label{Schroudinger}
\end{equation}%
has the following square integrable solution (Gaussian wave packet):%
\begin{equation}
\psi _{0}\left( x,t\right) =e^{i\left( \alpha \left( t\right) x^{2}+\delta
\left( t\right) x+\kappa \left( t\right) +\gamma \left( t\right) \right) }%
\sqrt{\frac{\beta \left( t\right) }{\sqrt{\pi }}}\ e^{-\left( \beta \left(
t\right) x+\varepsilon \left( t\right) \right) ^{2}/2},  \label{WaveFunction}
\end{equation}%
where%
\begin{eqnarray}
&&\alpha \left( t\right) =\frac{\alpha _{0}\cos 2t+\sin 2t\ \left( \beta
_{0}^{4}+4\alpha _{0}^{2}-1\right) /4}{\beta _{0}^{4}\sin ^{2}t+\left(
2\alpha _{0}\sin t+\cos t\right) ^{2}},  \label{hhA} \\
&&\beta \left( t\right) =\frac{\beta _{0}}{\sqrt{\beta _{0}^{4}\sin
^{2}t+\left( 2\alpha _{0}\sin t+\cos t\right) ^{2}}},  \label{hhB} \\
&&\gamma \left( t\right) =\gamma _{0}-\frac{1}{2}\arctan \frac{\beta
_{0}^{2}\tan t}{1+2\alpha _{0}\tan t},  \label{hhG} \\
&&\delta \left( t\right) =\frac{\delta _{0}\left( 2\alpha _{0}\sin t+\cos
t\right) +\varepsilon _{0}\beta _{0}^{3}\sin t}{\beta _{0}^{4}\sin
^{2}t+\left( 2\alpha _{0}\sin t+\cos t\right) ^{2}},  \label{hhD} \\
&&\varepsilon \left( t\right) =\frac{\varepsilon _{0}\left( 2\alpha _{0}\sin
t+\cos t\right) -\beta _{0}\delta _{0}\sin t}{\sqrt{\beta _{0}^{4}\sin
^{2}t+\left( 2\alpha _{0}\sin t+\cos t\right) ^{2}}},  \label{hhE} \\
&&\kappa \left( t\right) =\kappa _{0}+\sin ^{2}t\ \frac{\varepsilon
_{0}\beta _{0}^{2}\left( \alpha _{0}\varepsilon _{0}-\beta _{0}\delta
_{0}\right) -\alpha _{0}\delta _{0}^{2}}{\beta _{0}^{4}\sin ^{2}t+\left(
2\alpha _{0}\sin t+\cos t\right) ^{2}}  \label{hhK} \\
&&\qquad \quad +\frac{1}{4}\sin 2t\ \frac{\varepsilon _{0}^{2}\beta
_{0}^{2}-\delta _{0}^{2}}{\beta _{0}^{4}\sin ^{2}t+\left( 2\alpha _{0}\sin
t+\cos t\right) ^{2}}  \notag
\end{eqnarray}%
($\alpha _{0},$ $\beta _{0}\neq 0,$ $\gamma _{0},$ $\delta _{0},$ $%
\varepsilon _{0},$ $\kappa _{0}$ are real-valued initial data of the
corresponding Ermakov-type system; a complex form of equations (\ref{hhA})--(%
\ref{hhK}) is provided in Appendix~A and the invariants are given by (\ref%
{Invs1})--(\ref{Invs2}); in what follows one may choose $\gamma _{0}=\kappa
_{0}=0$). This quantum state can be thought of as a special case of a
`nonclassical' oscillator solution originally found by Marhic \cite{Marhic78}%
. The latter has been recently derived in a unified approach to generalized
harmonic oscillators (see, for example, \cite{Cor-Sot:Lop:Sua:Sus}, \cite%
{Cor-Sot:Sua:SusInv}, \cite{Lan:Lop:Sus}, \cite{LopSusVegaHarm}, \cite%
{SuazoSuslovSol} and the references therein). These solutions can be
verified by a direct substitution with the aid of \textsl{Mathematica}
computer algebra system \cite{Kouchan11} (see also \cite{KouchanZeilberger10}%
), \cite{LopSusVegaHarm}, and \cite{Lop:Sus:VegaMath}. (In retrospect, the\
simplest special case $\beta _{0}=\pm 1$ and $\alpha _{0}=\gamma _{0}=\delta
_{0}=\varepsilon _{0}=\kappa _{0}=0$ is the ground oscillator state \cite%
{Flu}, \cite{Gold:Krivch}, \cite{La:Lif}, \cite{Merz}, \cite%
{SchroedingerOscillator}, \cite{Schroedinger}. For the coherent states \cite%
{Schroedinger}, $\alpha _{0}=0$ and $\beta _{0}=\pm 1,$ while a more general
wave packet with $\alpha _{0}=0$ was discussed in \cite{Husimi53}, \cite%
{HusimiOtuka53}. Derivation of these formulas can be found in Refs.~\cite%
{Lop:Sus:VegaGroup}, \cite{LopSusVegaHarm}, and \cite{Marhic78}. An analog
of Berry's phase is evaluated in Refs.~\cite{Suslov12}, \cite{SuslovMath}.)

The \textquotedblleft dynamic harmonic oscillator ground
state\textquotedblright\ (\ref{WaveFunction})--(\ref{hhK}) is the
eigenfunction,%
\begin{equation}
E\left( t\right) \psi _{0}\left( x,t\right) =\frac{1}{2}\psi _{0}\left(
x,t\right) ,  \label{EigenValueProblem}
\end{equation}%
of the time-dependent dynamical invariant,%
\begin{eqnarray}
&&E\left( t\right) =\frac{1}{2}\left[ \frac{\left( p-2\alpha x-\delta
\right) ^{2}}{\beta ^{2}}+\left( \beta x+\varepsilon \right) ^{2}\right]
\label{QuadraticInvariant} \\
&&\qquad \ =\frac{1}{2}\left[ \widehat{a}\left( t\right) \widehat{a}%
^{\dagger }\left( t\right) +\widehat{a}^{\dagger }\left( t\right) \widehat{a}%
\left( t\right) \right] ,\quad \frac{d}{dt}\langle E\rangle =0,  \notag
\end{eqnarray}%
with a familiar operator identity:%
\begin{equation}
\frac{\partial E}{\partial t}+i^{-1}\left[ E,H\right] =0,\qquad H=\frac{1}{2}%
\left( p^{2}+x^{2}\right) .  \label{InvariantDer}
\end{equation}%
The time-dependent annihilation $\widehat{a}\left( t\right) $ and creation $%
\widehat{a}^{\dagger }\left( t\right) $ operators are given by the following
Bogoliubov-type transformation:%
\begin{eqnarray}
&&\widehat{a}\left( t\right) =\frac{1}{\sqrt{2}}\left( \beta x+\varepsilon +i%
\frac{p-2\alpha x-\delta }{\beta }\right) ,  \label{a(t)} \\
&&\widehat{a}^{\dagger }\left( t\right) =\frac{1}{\sqrt{2}}\left( \beta
x+\varepsilon -i\frac{p-2\alpha x-\delta }{\beta }\right)  \notag
\end{eqnarray}%
where $p=i^{-1}\partial /\partial x,$ in terms of solutions (\ref{hhA})--(%
\ref{hhK}) \cite{LopSusVegaHarm}. They satisfy the canonical commutation
relation,%
\begin{equation}
\widehat{a}\left( t\right) \widehat{a}^{\dagger }\left( t\right) -\widehat{a}%
^{\dagger }\left( t\right) \widehat{a}\left( t\right) =1,
\label{commutatora(t)across(t)}
\end{equation}%
and the spectrum of invariant $E$ can be obtained by using the
Heisenberg--Weyl algebra (a \textquotedblleft second
quantization\textquotedblright , the Fock states \cite{Fock32-2}, \cite%
{Fock34-3}, \cite{PerelomovCSBook}). In particular,%
\begin{equation}
\widehat{a}\left( t\right) \Psi _{0}\left( x,t\right) =0,\quad \psi
_{0}\left( x,t\right) =e^{i\gamma \left( t\right) }\ \Psi _{0}\left(
x,t\right) ,  \label{annandcratoperactions}
\end{equation}%
for the corresponding \textquotedblleft vacuum state\textquotedblright .

This form of quadratic dynamical invariant and creation and annihilation
operators for the generalized harmonic oscillators have been obtained in 
\cite{SanSusVin} (see also \cite{Cor-Sot:Sua:SusInv}, \cite{Dod:Man79}, \cite%
{Dodonov:Man'koFIAN87}, \cite{Suslov10} and the references therein for
important special cases). An application to the electro\-magnetic-field
quantization is discussed in \cite{Kr:Sus12} (see also section~8).

The maximum kinematical invariance groups of the free particle and harmonic
oscillator were introduced in \cite{AndersonPlus72}, \cite{AndersonII72}, 
\cite{Hagen72}, \cite{JACKIW80}, \cite{Niederer72}, and \cite{Niederer73}
(see also \cite{BoySharpWint}, \cite{KalninsMiller74}, \cite{Miller77}, \cite%
{Rosen76}, \cite{SuazoSusVega10}, \cite{SuazoSusVega11} and the references
therein). We use connections with the Ermakov-type system \cite%
{Lop:Sus:VegaGroup}, \cite{LopSusVegaHarm} (see \cite{Ermakov}, \cite%
{Leach:Andrio08} and the references therein regarding the Ermakov equation).
A general procedure of obtaining new solutions by using enveloping algebra
of generators of the Heisenberg--Weyl group is described in \cite%
{Dodonov:Man'koFIAN87} (see also \cite{Bag:Bel:Ter83}, \cite%
{Belov:Karavaev1987}, \cite{Dod:Man79}, \cite{Garraway00}, \cite{Marhic78}
regarding the corresponding wavefunctions).

\section{The Uncertainty Relation and Squeezing}

A quantum state is said to be \textquotedblleft squeezed\textquotedblright\
if its oscillating variances $\langle \left( \Delta p\right) ^{2}\rangle $
and $\langle \left( \Delta x\right) ^{2}\rangle $ become smaller than the
variances of the \textquotedblleft static\textquotedblright\ vacuum state $%
\langle \left( \Delta p\right) ^{2}\rangle =\langle \left( \Delta x\right)
^{2}\rangle =1/2$ (with $\hbar =1$). For the harmonic oscillator, the
product of the variances attains a minimum value only at the instances when
one variance is a minimum and the other is a maximum. If the minimum value
of the product is equal to $1/4,$ then the state is called a
minimum-uncertainty squeezed state (see, for example, \cite{DutraQED}, \cite%
{Henry:Glotzer88}, \cite{Kimetal89}, \cite{Shchukinetal09}, \cite{Stoler70}, 
\cite{Stoler71}, \cite{WallsMilburn08}, and \cite{Yuen76}). This property
can be easily verified for solution (\ref{WaveFunction}).

According to the transform (\ref{a(t)}), the corresponding expectation
values oscillate sinusoidally in time%
\begin{eqnarray}
&&\langle x\rangle =-\frac{1}{\beta _{0}}\left[ \left( 2\alpha
_{0}\varepsilon _{0}-\beta _{0}\delta _{0}\right) \sin t+\varepsilon
_{0}\cos t\right] ,\quad \dfrac{d}{dt}\langle x\rangle =\langle p\rangle ,
\label{<x>} \\
&&\langle p\rangle =-\frac{1}{\beta _{0}}\left[ \left( 2\alpha
_{0}\varepsilon _{0}-\beta _{0}\delta _{0}\right) \cos t-\varepsilon
_{0}\sin t\right] ,\quad \dfrac{d}{dt}\langle p\rangle =-\langle x\rangle
\label{<p>}
\end{eqnarray}%
with the initial data $\left. \langle x\rangle \right\vert
_{t=0}=-\varepsilon _{0}/\beta _{0}$ and $\left. \langle p\rangle
\right\vert _{t=0}=-\left( 2\alpha _{0}\varepsilon _{0}-\beta _{0}\delta
_{0}\right) /\beta _{0}.$ This provides a connection of these parameters
with the Ehrenfest theorem \cite{Ehrenfest}, \cite{HeisenbergQM}, \cite%
{WeinbergQM}.

The expectation values $\langle x\rangle $ and $\langle p\rangle $ satisfy
the classical equation for harmonic motion, $y^{\prime \prime }+y=0,$ with
the total \textquotedblleft classical mechanical energy\textquotedblright\
given by%
\begin{equation}
\frac{1}{2}\left[ \langle p\rangle ^{2}+\langle x\rangle ^{2}\right] =\frac{%
\left( 2\alpha _{0}\varepsilon _{0}-\beta _{0}\delta _{0}\right)
^{2}+\varepsilon _{0}^{2}}{2\beta _{0}^{2}}=\left. \frac{1}{2}\left[ \langle
p\rangle ^{2}+\langle x\rangle ^{2}\right] \right\vert _{t=0}.
\label{ClassMechEnergy}
\end{equation}%
For the standard deviations on solution (\ref{WaveFunction})--(\ref{hhK}),
one gets%
\begin{eqnarray}
&&\langle \left( \Delta p\right) ^{2}\rangle =\langle p^{2}\rangle -\langle
p\rangle ^{2}  \label{DeltaPG} \\
&&\quad =\dfrac{1+4\alpha _{0}^{2}+\beta _{0}^{4}+\left( 4\alpha
_{0}^{2}+\beta _{0}^{4}-1\right) \cos 2t-4\alpha _{0}\sin 2t}{4\beta _{0}^{2}%
},  \notag \\
&&\langle \left( \Delta x\right) ^{2}\rangle =\langle x^{2}\rangle -\langle
x\rangle ^{2}  \label{DeltaXG} \\
&&\quad =\dfrac{1+4\alpha _{0}^{2}+\beta _{0}^{4}-\left( 4\alpha
_{0}^{2}+\beta _{0}^{4}-1\right) \cos 2t+4\alpha _{0}\sin 2t}{4\beta _{0}^{2}%
},  \notag
\end{eqnarray}%
and%
\begin{eqnarray}
&&\langle \left( \Delta p\right) ^{2}\rangle \langle \left( \Delta x\right)
^{2}\rangle =\dfrac{1}{16\beta _{0}^{4}}  \label{HUR} \\
&&\quad \times \left[ \left( 1+4\alpha _{0}^{2}+\beta _{0}^{4}\right)
^{2}-\left( \left( 4\alpha _{0}^{2}+\beta _{0}^{4}-1\right) \cos 2t-4\alpha
_{0}\sin 2t\right) ^{2}\right] .  \notag
\end{eqnarray}%
Here,%
\begin{eqnarray}
&&\sigma _{p}=\langle \left( \Delta p\right) ^{2}\rangle =\frac{4\alpha
^{2}+\beta ^{4}}{2\beta ^{2}},\quad \sigma _{x}=\langle \left( \Delta
x\right) ^{2}\rangle =\frac{1}{2\beta ^{2}},  \label{Variences} \\
&&\quad \quad \qquad \sigma _{px}=\frac{1}{2}\langle \Delta p\Delta x+\Delta
x\Delta p\rangle =\frac{\alpha }{\beta ^{2}}  \notag
\end{eqnarray}%
with two invariants:%
\begin{equation}
\sigma _{p}+\sigma _{x}=\frac{4\alpha ^{2}+\beta ^{4}+1}{2\beta ^{2}}=\frac{%
4\alpha _{0}^{2}+\beta _{0}^{4}+1}{2\beta _{0}^{2}},\qquad \left\vert 
\begin{array}{cc}
\sigma _{p} & \sigma _{px} \\ 
\sigma _{px} & \sigma _{x}%
\end{array}%
\right\vert =\sigma _{p}\sigma _{x}-\sigma _{px}^{2}=\frac{1}{4}
\label{InvariantsSigma}
\end{equation}%
(More invariants are given by in (\ref{Invs1})--(\ref{Invs2}).) The Schr\"{o}%
dinger minimum-uncertainty states \cite{Schroedinger}, when $\langle \left(
\Delta p\right) ^{2}\rangle =\langle \left( \Delta x\right) ^{2}\rangle
=1/2, $ are defined by taking $\alpha _{0}=0$ and $\beta _{0}^{2}=1.$ For
the ground state solution, when $\alpha _{0}=\delta _{0}=\varepsilon _{0}=0$
and $\beta _{0}=\pm 1,$ one gets $\langle x\rangle =\langle p\rangle \equiv
0 $ and%
\begin{equation}
\langle \left( \Delta p\right) ^{2}\rangle =\langle \left( \Delta x\right)
^{2}\rangle =\frac{1}{2}  \label{Vacuum}
\end{equation}%
as presented in the textbooks \cite{Flu}, \cite{Gold:Krivch}, \cite%
{Guerr:Lop:Ald:Coss11}, \cite{Henry:Glotzer88}, \cite{La:Lif}, \cite{Merz}, 
\cite{PerelomovCSBook}.

By adding (\ref{ClassMechEnergy})--(\ref{DeltaXG}), we arrive at 
\begin{eqnarray}
&&\langle H\rangle =\frac{1}{2}\left[ \langle p^{2}\rangle +\langle
x^{2}\rangle \right]  \label{QuantumEnergy} \\
&&\ =\dfrac{1+4\alpha _{0}^{2}+\beta _{0}^{4}}{4\beta _{0}^{2}}+\frac{\left(
2\alpha _{0}\varepsilon _{0}-\beta _{0}\delta _{0}\right) ^{2}+\varepsilon
_{0}^{2}}{2\beta _{0}^{2}}\geq \frac{1}{2}  \notag
\end{eqnarray}%
for the total \textquotedblleft quantum mechanical energy\textquotedblright\
in terms of integrals of motion (the vacuum value $1/2$ occurs when $\beta
_{0}=\pm 1$ and $\alpha _{0}=\delta _{0}=\varepsilon _{0}=0).$ See also \cite%
{Breit:Schill:Mlyn97} and \cite{Dod:Man:Man94}.

Therefore, the upper and lower bound in the Heisenberg uncertainty relation
are given by%
\begin{equation}
\max \left[ \langle \left( \Delta p\right) ^{2}\rangle \langle \left( \Delta
x\right) ^{2}\rangle \right] =\frac{\left( 1+4\alpha _{0}^{2}+\beta
_{0}^{4}\right) ^{2}}{16\beta _{0}^{4}},\qquad \text{when\quad }\cot 2t=%
\frac{4\alpha _{0}}{4\alpha _{0}^{2}+\beta _{0}^{4}-1}  \label{MAX}
\end{equation}%
and%
\begin{equation}
\min \left[ \langle \left( \Delta p\right) ^{2}\rangle \langle \left( \Delta
x\right) ^{2}\rangle \right] =\frac{1}{4},\qquad \text{if\quad }\tan 2t=-%
\frac{4\alpha _{0}}{4\alpha _{0}^{2}+\beta _{0}^{4}-1}.  \label{min}
\end{equation}%
Our explicit formulas (\ref{DeltaPG})--(\ref{DeltaXG}) show that the product
of the variances attains the minimum value $1/4$ only at the instances that
one variance is a minimum and the other is a maximum as stated in \cite%
{Henry:Glotzer88}. Here, squeezing of one of the variances is explicitly
described. Indeed, 
\begin{equation}
\left( 4\alpha _{0}^{2}+\beta _{0}^{4}-1\right) \cos 2t-4\alpha _{0}\sin
2t=\pm \left( 4\alpha _{0}^{2}+\left( \beta _{0}^{2}+1\right) ^{2}\right)
^{1/2}\left( 4\alpha _{0}^{2}+\left( \beta _{0}^{2}-1\right) ^{2}\right)
^{1/2}\ ,  \label{sincos}
\end{equation}%
under the minimization condition (\ref{min}) and at the minimum%
\begin{eqnarray}
&&\langle \left( \Delta p\right) ^{2}\rangle =\dfrac{1}{4\beta _{0}^{2}}%
\left[ 1+4\alpha _{0}^{2}+\beta _{0}^{4}\pm \left( 4\alpha _{0}^{2}+\left(
\beta _{0}^{2}+1\right) ^{2}\right) ^{1/2}\left( 4\alpha _{0}^{2}+\left(
\beta _{0}^{2}-1\right) ^{2}\right) ^{1/2}\right] ,  \label{DP} \\
&&\langle \left( \Delta x\right) ^{2}\rangle =\dfrac{1}{4\beta _{0}^{2}}%
\left[ 1+4\alpha _{0}^{2}+\beta _{0}^{4}\mp \left( 4\alpha _{0}^{2}+\left(
\beta _{0}^{2}+1\right) ^{2}\right) ^{1/2}\left( 4\alpha _{0}^{2}+\left(
\beta _{0}^{2}-1\right) ^{2}\right) ^{1/2}\right]  \label{DX}
\end{eqnarray}%
for all real values of our parameters. At this instant the squeezing occur:%
\begin{equation*}
\langle \left( \Delta p\right) ^{2}\rangle >\frac{1}{2}\left( <\frac{1}{2}%
\right) ,\qquad \langle \left( \Delta x\right) ^{2}\rangle <\frac{1}{2}%
\left( >\frac{1}{2}\right)
\end{equation*}%
(for upper and lower signs, respectively). As a result, the
minimum-uncertainty squeezed states for the linear harmonic oscillator are
presented in closed form (\ref{hhA})--(\ref{hhK}) (see also \cite%
{Henry:Glotzer88} for numerical simulations). A natural generalization will
be discussed in the next section. The corresponding wavefunction in the
momentum representation can be derived by the (inverse) Fourier transform of
(\ref{WaveFunction}) and (\ref{hhA})--(\ref{hhK}) (see \cite{LopSusVegaHarm}
for more details). Experimentally observed time-oscillations of the velocity
variance \cite{Morinagaetal99} reveal certain damping, which can be explain
in models of quantum damped oscillators discussed in \cite%
{Cor-SotSuaSusDamped}, \cite{Cor-Sot:Sua:SusInv}, and \cite{Dod:Man79} (see
also the references therein).

\section{An Extension: the TCS States}

We construct an analog of the coherent states (generalized coherent, or the
TCS states in the terminology of Ref.~ \cite{Yuen76}) in a standard fashion%
\begin{eqnarray}
&&\psi \left( x,t\right) =e^{-\left\vert \zeta \right\vert
^{2}/2}\sum_{n=0}^{\infty }\psi _{n}\left( x,t\right) \ \frac{\zeta ^{n}}{%
\sqrt{n!}}  \label{TCS} \\
&&\ =e^{-\left\vert \eta \right\vert ^{2}/2}e^{i\gamma }\sum_{n=0}^{\infty
}\Psi _{n}\left( x,t\right) \ \frac{\eta ^{n}}{\sqrt{n!}},\quad \eta =\zeta
e^{2i\gamma },  \notag
\end{eqnarray}%
where $\zeta $ is an arbitrary complex parameter and the \textquotedblleft
dynamic\textquotedblright\ wavefunctions are given by equations (1.2) and
(1.16) of \cite{LopSusVegaHarm} reproduced here for the reader's
convenience: 
\begin{equation}
\psi _{n}\left( x,t\right) =e^{i\left( \alpha x^{2}+\delta x+\kappa \right)
+i\left( 2n+1\right) \gamma }\sqrt{\frac{\beta }{2^{n}n!\sqrt{\pi }}}\
e^{-\xi ^{2}/2}\ H_{n}\left( \xi \right) ,\qquad \xi =\beta x+\varepsilon
\label{WaveFunctionN}
\end{equation}%
(see also \cite{Dod:Man79} and \cite{Marhic78}), where $H_{n}\left( x\right) 
$ are the Hermite polynomials \cite{Ni:Su:Uv}. In the explicit form \cite%
{Schroedinger},%
\begin{eqnarray}
&&\psi \left( x,t\right) =\sqrt{\frac{\beta }{\sqrt{\pi }}}e^{-\left( \xi
^{2}+\left\vert \eta \right\vert ^{2}\right) /2}e^{i\left( \alpha
x^{2}+\delta x+\kappa +\gamma \right) }\sum_{n=0}^{\infty }\left( \frac{\eta 
}{\sqrt{2}}\right) ^{n}\frac{H_{n}\left( \xi \right) }{n!}  \label{Hermite}
\\
&&\qquad \quad \ =\sqrt{\frac{\beta }{\sqrt{\pi }}}e^{\left( \eta
^{2}-\left\vert \eta \right\vert ^{2}\right) /2}e^{i\left( \alpha
x^{2}+\delta x+\kappa +\gamma \right) }e^{-\left( \xi -\sqrt{2}\eta \right)
^{2}/2},  \notag
\end{eqnarray}%
and the eigenvalue problem is given by \cite{Yuen76}:%
\begin{equation}
\widehat{a}\left( t\right) \psi \left( x,t\right) =\eta \psi \left(
x,t\right) .  \label{eigeneta}
\end{equation}%
An elementary calculation shows that on these\ \textquotedblleft dynamic
coherent states\textquotedblright , 
\begin{equation}
\langle x\rangle =\frac{1}{\beta \sqrt{2}}\left( \eta +\eta ^{\ast }\right) -%
\frac{\varepsilon }{\beta },\qquad \quad \left. \langle x\rangle \right\vert
_{t=0}=\frac{\sqrt{2}}{\beta _{0}}\left\vert \zeta \right\vert \cos \left(
2\left( \gamma _{0}+\phi \right) \right) -\frac{\varepsilon _{0}}{\beta _{0}}%
,  \label{AverageX}
\end{equation}%
and%
\begin{eqnarray}
&&\langle p\rangle =\frac{\beta }{i\sqrt{2}}\left( \eta -\eta ^{\ast
}\right) +\frac{\alpha \sqrt{2}}{\beta }\left( \eta +\eta ^{\ast }\right)
+\left( \delta -\frac{2\alpha \varepsilon }{\beta }\right) ,
\label{AveragePP} \\
&&\left. \langle p\rangle \right\vert _{t=0}=\beta _{0}\sqrt{2}\left\vert
\zeta \right\vert \sin \left( 2\left( \gamma _{0}+\phi \right) \right)
+2^{3/2}\frac{\alpha _{0}}{\beta _{0}}\left\vert \zeta \right\vert \cos
\left( 2\left( \gamma _{0}+\phi \right) \right) +\left( \delta _{0}-\frac{%
2\alpha _{0}\varepsilon _{0}}{\beta _{0}}\right) ,  \notag
\end{eqnarray}%
if $\zeta =\left\vert \zeta \right\vert e^{2i\phi }.$ Moreover, a direct 
\textsl{Mathematica} verification shows that these expectation values
satisfy the required classical equation for simple harmonic motion.

A similar calculation reveals that the corresponding oscillating variances $%
\langle \left( \Delta p\right) ^{2}\rangle $ and $\langle \left( \Delta
x\right) ^{2}\rangle $ coincide with those for the \textquotedblleft dynamic
vacuum states\textquotedblright\ given by (\ref{DeltaPG})--(\ref{DeltaXG}).
The \textquotedblleft dynamic coherent states\textquotedblright\ (\ref%
{Hermite}) are also the minimum-uncertainty squeezed states but they are not
eigenfunctions of the time-dependent dynamic invariant (\ref%
{QuadraticInvariant}) when $\eta \neq 0.$

The Wigner function \cite{HilletyetalWigner84}, \cite{LeonardPaul95}, \cite%
{Schleich01}, \cite{Schradeetal95}, \cite{Wigner32},%
\begin{equation}
W\left( x,p\right) =\frac{1}{2\pi }\int_{-\infty }^{\infty }\psi ^{\ast
}\left( x+y/2\right) \psi \left( x-y/2\right) e^{ipy}\ dy,
\label{WignerDefinition}
\end{equation}%
for the TCS states (\ref{Hermite}) is given by%
\begin{equation}
W\left( x,p\right) =\frac{1}{\pi }\exp \left[ -\left( P+i\frac{\eta -\eta
^{\ast }}{\sqrt{2}}\right) ^{2}-\left( Q-\frac{\eta +\eta ^{\ast }}{\sqrt{2}}%
\right) ^{2}\right] ,  \label{WignerTCS}
\end{equation}%
where%
\begin{equation}
P=\frac{p-2\alpha x-\delta }{\beta },\qquad Q=\beta x+\varepsilon .
\label{GeneralizedPQ}
\end{equation}%
In view of (\ref{AverageX})--(\ref{AveragePP}), we arrive at the following
expression of the Wigner function:%
\begin{equation}
W\left( x,p\right) =\frac{1}{\pi }\exp \left[ -\frac{\left( p-\langle
p\rangle \right) ^{2}}{\beta ^{2}}+\frac{4\alpha }{\beta ^{2}}\left(
p-\langle p\rangle \right) \left( x-\langle x\rangle \right) -\frac{4\alpha
^{2}+\beta ^{4}}{\beta ^{2}}\left( x-\langle x\rangle \right) ^{2}\right] ,
\label{WignerErmakov}
\end{equation}%
in terms of the classical trajectories $\langle x\rangle $ and $\langle
p\rangle $ and the solutions of Ermakov-type system (\ref{hhA})--(\ref{hhB}%
). Taking into account the time-dependent variances (\ref{Variences}), one
gets \cite{Dod:Kur:Man80}, \cite{Dod:Man:Man94}, \cite{Schradeetal95}, \cite%
{Stenholm86}: 
\begin{equation}
W\left( x,p\right) =\frac{1}{\pi }\exp \left[ -2\left( \sigma _{x}\left(
p-\langle p\rangle \right) ^{2}-2\sigma _{px}\left( p-\langle p\rangle
\right) \left( x-\langle x\rangle \right) +\sigma _{p}\left( x-\langle
x\rangle \right) ^{2}\right) \right] ,  \label{WignerClassics}
\end{equation}%
where $\sigma _{p},$ $\sigma _{x},$ and $\sigma _{px}$ are given by (\ref%
{Variences}). Then%
\begin{equation}
W\left( x,p;t\right) =W\left( x\cos t-p\sin t,x\sin t+p\cos t;t=0\right)
\label{WignerRotate}
\end{equation}%
by a direct calculation --- the graph of Wigner function rotates in the
phase plane without changing its shape \cite{Stenholm86}. (In a traditional
approach, the quantum Liouville equation of motion for Wigner function of
the corresponding quadratic system is used in order to determine this time
evolution \cite{Schleich01}. We have obtained the same result directly from
the wavefunctions; see also \cite{Schradeetal95}.) Some \textsl{Mathematica}
animations can be found in Ref.~\cite{KrSusVegaWignerMath}. Reconstruction
of the original wavefunction from the Wigner quasidistribution is discussed
in Refs.~\cite{LeavenSalaMayato01}, \cite{Takabayasi54}.~(See also Ref.~\cite%
{Klauder60} for a detailed discussion of fundamental limitations on
simultaneous range-velocity determination in radar systems with the aid of a
quantum mechanical analog to the Wigner distribution function.)

\section{The Moyal Functions}

The total energy of a \textquotedblleft dynamic harmonic
state\textquotedblright\ (\ref{WaveFunctionN}) can be presented as 
\begin{equation}
\langle H\rangle =\frac{1}{2}\left[ \langle p^{2}\rangle +\langle
x^{2}\rangle \right] =\left( n+\frac{1}{2}\right) \dfrac{1+4\alpha
_{0}^{2}+\beta _{0}^{4}}{2\beta _{0}^{2}}+\frac{\left( 2\alpha
_{0}\varepsilon _{0}-\beta _{0}\delta _{0}\right) ^{2}+\varepsilon _{0}^{2}}{%
2\beta _{0}^{2}}  \label{Energy}
\end{equation}%
by (A.3)--(A.5) of Ref.\ \cite{LopSusVegaHarm}.

The Moyal functions \cite{Moyal47} for the \textquotedblleft dynamic
harmonic states\textquotedblright\ (\ref{WaveFunctionN}): 
\begin{equation}
W_{mn}\left( x,p,t\right) =\frac{1}{2\pi }\int_{-\infty }^{\infty }\psi
_{m}^{\ast }\left( x+y/2,t\right) \psi _{n}\left( x-y/2,t\right) e^{ipy}\ dy
\label{MoyalFunction}
\end{equation}%
can be evaluated in terms of Laguerre and Charlier polynomials in the
standard way \cite{Klauder60}, \cite{Ni:Su:Uv}, \cite{Schleich01}, \cite%
{Schradeetal95}:%
\begin{eqnarray}
W_{mn}\left( x,p,t\right) &=&\frac{\left( -1\right) ^{m}e^{2i\left(
n-m\right) \gamma }}{\pi }e^{-Q^{2}-P^{2}}2^{\left( m-n\right) /2}\sqrt{%
\frac{m!}{n!}}  \label{MoyalFunctionLaguerre} \\
&&\times \left( Q-iP\right) ^{n-m}L_{m}^{n-m}\left( 2\left(
Q^{2}+P^{2}\right) \right)  \notag
\end{eqnarray}%
in the notation (\ref{GeneralizedPQ}). Once again, the time evolution of the
corresponding Wigner function $W_{nn}\left( x,p,t\right) $ is defined by
equation (\ref{WignerRotate}).

In the case of an arbitrary linear combination,%
\begin{equation}
\psi \left( x,t\right) =\sum_{m}c_{m}\psi _{m}\left( x,t\right) ,
\label{LinWave}
\end{equation}%
the Wigner function can be obtain as a double sum of Moyal's functions:%
\begin{equation}
W\left( x,p,t\right) =\sum_{m,n}c_{m}^{\ast }c_{n}W_{mn}\left( x,p,t\right) .
\label{LinWigner}
\end{equation}%
A coherent superposition of two Fock states with $n=0$ and $n=1$ was
experimentally realized in \cite{Morinagaetal99}. Moreover, the state of the
electromagnetic field can be chosen anywhere between the single-photon and
squeezed state in Ref.~\cite{JainetalLv10}.

\section{Eigenfunction Expansions}

Experimentally observed statistics for various squeezed states of photons
and ions in a box \cite{Breit:Schill:Mlyn97}, \cite{HarocheRaimond06}, \cite%
{LeibfriedetalWineland03}, \cite{LvRay09}, \cite{Meekhofetal96}, \cite%
{SchillerBreitenbachetal96} can be naturally explained in terms of explicit
developments with respect to the Fock states. For a linear harmonic
oscillator in coordinate representation, we consider the corresponding
wavefunctions and use known expansions in Hermite polynomials \cite{Lan:Sus}%
, \cite{Lop:Sus}, \cite{Ni:Su:Uv}. Group-theoretical properties are
discussed, for example, in Refs.~\cite{Dodonov:Man'koFIAN87}, \cite%
{Kimetal89}, \cite{Ni:Su:Uv}, \cite{PerelomovCSBook}.

\subsection{Familiar Expansions}

For the stationary harmonic oscillator wavefunctions,%
\begin{equation}
\Psi _{n}\left( x\right) =\frac{e^{-x^{2}/2}}{\sqrt{2^{n}n!\sqrt{\pi }}}\
H_{n}\left( x\right) ,  \label{HarmonicWaveFunctions}
\end{equation}%
there are two well known expansions:%
\begin{equation}
e^{i\left( \Gamma +Bx\right) }\Psi _{n}\left( x+A\right) =\sum_{m=0}^{\infty
}T_{mn}\left( A,B,\Gamma \right) \ \Psi _{m}\left( x\right) ,
\label{HeisengergGroup}
\end{equation}%
where 
\begin{eqnarray}
T_{mn}\left( A,B,\Gamma \right) &=&\int_{-\infty }^{\infty }\Psi _{m}^{\ast
}\left( x\right) e^{i\left( \Gamma +Bx\right) }\Psi _{n}\left( x+A\right) \
dx  \label{HeisenbergGroupMatrix} \\
&=&\frac{i^{m-n}}{\sqrt{m!n!}}\ e^{i\left( \Gamma -AB/2\right) }\ e^{-\nu
/2}\ \left( \frac{iA+B}{\sqrt{2}}\right) ^{m}\left( \frac{iA-B}{\sqrt{2}}%
\right) ^{n}  \notag \\
&&\times \ _{2}F_{0}\left( -n,\ -m;\ -\frac{1}{\nu }\right)  \notag
\end{eqnarray}%
with $\nu =\left( A^{2}+B^{2}\right) /2$ (see, for example, \cite{Lop:Sus}, 
\cite{Ni:Su:Uv} for relations with the Heisenberg--Weyl group, Charlier
polynomials, and Poisson distribution) and%
\begin{equation}
e^{i\alpha x^{2}}\Psi _{n}\left( \beta x\right) =\sum_{m=0}^{\infty
}M_{mn}\left( \alpha ,\beta \right) \ \Psi _{m}\left( x\right) .
\label{SU(1,1)}
\end{equation}%
By the orthogonality,%
\begin{equation}
M_{mn}\left( \alpha ,\beta \right) =\int_{-\infty }^{\infty }\Psi _{m}^{\ast
}\left( x\right) e^{i\alpha x^{2}}\Psi _{n}\left( \beta x\right) \ dx,
\label{SU(1,1)Matrix}
\end{equation}%
and one can use the integral evaluated by Bailey:%
\begin{eqnarray}
&&\int_{-\infty }^{\infty }e^{-\lambda ^{2}x^{2}}H_{m}\left( ax\right)
H_{n}\left( bx\right) \ dx  \label{Bai} \\
&&\quad =\frac{2^{m+n}}{\lambda ^{m+n+1}}\Gamma \left( \frac{m+n+1}{2}%
\right) \left( a^{2}-\lambda ^{2}\right) ^{m/2}\left( b^{2}-\lambda
^{2}\right) ^{n/2}  \notag \\
&&\qquad \times ~_{2}F_{1}\left( 
\begin{array}{c}
-m,\quad -n \\ 
\dfrac{1}{2}\left( 1-m-n\right)%
\end{array}%
;\dfrac{1}{2}\left( 1-\frac{ab}{\sqrt{\left( a^{2}-\lambda ^{2}\right)
\left( b^{2}-\lambda ^{2}\right) }}\right) \right) ,  \notag
\end{eqnarray}%
$\func{Re}\lambda ^{2}>0,$ if $m+n$ is even; the integral vanishes by
symmetry if $m+n$ is odd; see \cite{Bailey48}, \cite{Lord49} and the
references therein for earlier works on these integrals, some of their
special cases and extensions. As a result,%
\begin{eqnarray}
&&M_{mn}\left( \alpha ,\beta \right) =i^{n}\sqrt{\frac{2^{m+n}}{m!n!\pi }}\
\Gamma \left( \frac{m+n+1}{2}\right)  \label{MartixSU(1,1)Hyper} \\
&&\quad \times \frac{\left( \dfrac{1-\beta ^{2}}{2}+i\alpha \right)
^{m/2}\left( \dfrac{1-\beta ^{2}}{2}-i\alpha \right) ^{n/2}}{\left( \dfrac{%
1+\beta ^{2}}{2}-i\alpha \right) ^{\left( m+n+1\right) /2}}  \notag \\
&&\quad \times ~_{2}F_{1}\left( 
\begin{array}{c}
-m,\quad -n \\ 
\dfrac{1}{2}\left( 1-m-n\right)%
\end{array}%
;\dfrac{1}{2}\left( 1\pm \frac{2i\beta }{\sqrt{4\alpha ^{2}+\left( \beta
^{2}-1\right) ^{2}}}\right) \right) .  \notag
\end{eqnarray}%
The terminating hypergeometric function can be transformed as follows%
\begin{eqnarray}
&&_{2}F_{1}\left( 
\begin{array}{c}
-k,\quad -n \\ 
\dfrac{1}{2}\left( 1-k-n\right)%
\end{array}%
;\quad \frac{1}{2}\left( 1+i\zeta \right) \right) \medskip  \label{spc21a} \\
&&\ =\left\{ 
\begin{array}{ll}
\dfrac{\left( 1/2\right) _{r}\left( 1/2\right) _{s}}{\left( 1/2\right) _{r+s}%
}\ _{2}F_{1}\left( 
\begin{array}{c}
-r,\quad -s\medskip \\ 
1/2%
\end{array}%
;\quad -\zeta ^{2}\right) ,\medskip & \text{if }k=2r,\ n=2s,\medskip \\ 
-\dfrac{\left( 3/2\right) _{r}\left( 3/2\right) _{s}}{\left( 3/2\right)
_{r+s}}\ i\zeta ~_{2}F_{1}\left( 
\begin{array}{c}
-r,\quad -s\medskip \\ 
3/2%
\end{array}%
;\quad -\zeta ^{2}\right) , & \text{if }k=2r+1,\ n=2s+1.%
\end{array}%
\right.  \notag
\end{eqnarray}%
It is valid in the entire complex plane; the details are given in Appendix~B
of \cite{Lan:Sus}. The latter transformation completes evaluation of the
Bailey integral (\ref{Bai}) and the matrix elements (\ref{MartixSU(1,1)Hyper}%
) in terms of the hypergeometric functions. (Relations with the group $%
SU\left( 1,1\right) ,$ Meixner polynomials \cite{Ni:Su:Uv}, and with two
special cases of the negative binomial, or Pascal, distribution \cite%
{Lan:Sus} are discussed elsewhere.)

\subsection{Probability Amplitudes}

Expansions (\ref{HeisengergGroup}) and (\ref{SU(1,1)}) results in%
\begin{equation}
\psi _{n}\left( x,t\right) =e^{i\left( 2n+1\right) \left( \gamma -\gamma
_{0}\right) }\sqrt{\beta }\sum_{m=0}^{\infty }C_{mn}\left( t\right) \ \Psi
_{m}\left( x\right) ,  \label{ExpansionFinal}
\end{equation}%
where%
\begin{eqnarray}
C_{mn}\left( t\right) &=&\sum_{k=0}^{\infty }M_{mk}\left( \alpha ,\beta
\right) \ T_{kn}\left( \varepsilon ,\frac{\delta }{\beta },\kappa \right)
\label{ExpansionCoefficients} \\
&=&\sum_{k=0}^{\infty }T_{mk}\left( \frac{\varepsilon }{\beta },\delta -%
\frac{2\alpha \varepsilon }{\beta },\kappa -\frac{\alpha \varepsilon ^{2}}{%
\beta ^{2}}\right) \ M_{kn}\left( \alpha ,\beta \right) .  \notag
\end{eqnarray}%
The invariants are%
\begin{eqnarray}
&&\frac{4\alpha ^{2}+\beta ^{4}+1}{2\beta ^{2}}=\frac{4\alpha _{0}^{2}+\beta
_{0}^{4}+1}{2\beta _{0}^{2}},\qquad \kappa -\frac{\delta \varepsilon }{%
2\beta }=\kappa _{0}-\frac{\delta _{0}\varepsilon _{0}}{2\beta _{0}},
\label{Invs1} \\
&&\varepsilon ^{2}+\frac{\delta ^{2}}{\beta ^{2}}=\varepsilon _{0}^{2}+\frac{%
\delta _{0}^{2}}{\beta _{0}^{2}},\qquad \frac{\varepsilon ^{2}}{\beta ^{2}}%
+\left( \delta -\frac{2\alpha \varepsilon }{\beta }\right) ^{2}=\frac{%
\varepsilon _{0}^{2}}{\beta _{0}^{2}}+\left( \delta _{0}-\frac{2\alpha
_{0}\varepsilon _{0}}{\beta _{0}}\right) ^{2}  \label{Invs2}
\end{eqnarray}%
by a direct calculation. Another useful identity is given by%
\begin{equation}
\frac{4\alpha ^{2}+\beta ^{4}+1}{2\beta ^{2}}\pm 1=\frac{4\alpha ^{2}+\left(
\beta ^{2}\pm 1\right) ^{2}}{2\beta ^{2}}=\sigma _{p}+\sigma _{x}\pm 1.
\label{Invs3}
\end{equation}%
Thus all arguments of the hypergeometric functions in (\ref%
{ExpansionCoefficients}) are constants. Moreover, the time-dependencies of
the matrix elements are given by complex phase factors only:%
\begin{eqnarray}
&&T_{mn}\left( \varepsilon ,\frac{\delta }{\beta },\kappa \right)
=e^{2i\left( m-n\right) \left( \gamma -\gamma _{0}\right) }\ T_{mn}\left(
\varepsilon _{0},\frac{\delta _{0}}{\beta _{0}},\kappa _{0}\right) ,
\label{Mat1} \\
&&T_{mn}\left( \frac{\varepsilon }{\beta },\delta -\frac{2\alpha \varepsilon 
}{\beta },\kappa -\frac{\alpha \varepsilon ^{2}}{\beta ^{2}}\right)
=e^{i\left( n-m\right) t}\ T_{mn}\left( \frac{\varepsilon _{0}}{\beta _{0}}%
,\delta _{0}-\frac{2\alpha _{0}\varepsilon _{0}}{\beta _{0}},\kappa _{0}-%
\frac{\alpha _{0}\varepsilon _{0}^{2}}{\beta _{0}^{2}}\right)  \label{Mat2}
\end{eqnarray}%
and%
\begin{equation}
M_{mn}\left( \alpha ,\beta \right) =e^{-i\left( 2m+1\right) \left( \gamma
-\gamma _{0}\right) }e^{-i\left( n+1/2\right) t}\sqrt{\frac{\beta _{0}}{%
\beta }}\ M_{mn}\left( \alpha _{0},\beta _{0}\right)  \label{Mat3}
\end{equation}%
in view of the following identities%
\begin{eqnarray}
\frac{\delta }{\beta }+i\varepsilon &=&\left( \frac{\delta _{0}}{\beta _{0}}%
+i\varepsilon _{0}\right) e^{2i\left( \gamma -\gamma _{0}\right) },
\label{Id1} \\
\delta -\frac{2\alpha \varepsilon }{\beta }+i\frac{\varepsilon }{\beta }
&=&\left( \delta _{0}-\frac{2\alpha _{0}\varepsilon _{0}}{\beta _{0}}+i\frac{%
\varepsilon _{0}}{\beta _{0}}\right) e^{-it},  \label{Id2} \\
\frac{1-\beta ^{2}}{2}+i\alpha &=&e^{-it}\left( \frac{1-\beta _{0}^{2}}{2}%
+i\alpha _{0}\right) /\left( 2\alpha _{0}\sin t+\cos t+i\beta _{0}^{2}\sin
t\right) ,  \label{Id3} \\
\frac{1+\beta ^{2}}{2}-i\alpha &=&e^{it}\left( \frac{1+\beta _{0}^{2}}{2}%
-i\alpha _{0}\right) /\left( 2\alpha _{0}\sin t+\cos t+i\beta _{0}^{2}\sin
t\right)  \label{Id4}
\end{eqnarray}%
and some of their complex conjugates (see also Appendix~A for a complex
parametrization of the Schr\"{o}dinger group).

Finally, the eigenfunction expansion takes the form%
\begin{equation}
\psi _{n}\left( x,t\right) =\sqrt{\beta _{0}}\sum_{m=0}^{\infty }c_{mn}\
e^{-i\left( m+1/2\right) t}\ \Psi _{m}\left( x\right) ,
\label{ExpansionFinalIndependent}
\end{equation}%
where the time-independent coefficients are explicitly given by%
\begin{eqnarray}
c_{mn} &=&\sum_{k=0}^{\infty }M_{mk}\left( \alpha _{0},\beta _{0}\right) \
T_{kn}\left( \varepsilon _{0},\frac{\delta _{0}}{\beta _{0}},\kappa
_{0}\right)  \label{ExpansionCoeffs} \\
&=&\sum_{k=0}^{\infty }T_{mk}\left( \frac{\varepsilon _{0}}{\beta _{0}}%
,\delta _{0}-\frac{2\alpha _{0}\varepsilon _{0}}{\beta _{0}},\kappa _{0}-%
\frac{\alpha _{0}\varepsilon _{0}^{2}}{\beta _{0}^{2}}\right) \ M_{kn}\left(
\alpha _{0},\beta _{0}\right)  \notag
\end{eqnarray}%
in terms of the initial data/integrals of motion (of the corresponding
Ermakov-type system). Thus the total probability amplitude is connected to
the product of two infinite matrices related to the Poisson and Pascal
distributions.

Moreover, a combination of (\ref{TCS}) and (\ref{ExpansionFinalIndependent})
gives the eigenfunction expansion of the TCS states. It is worth noting also
that our expansion (\ref{ExpansionFinalIndependent}) gives an independent
verification of the fact that the \textquotedblleft
missing\textquotedblright \footnote{%
omitted in \textquotedblleft The Bible of Theoretical
Physics\textquotedblright\ \cite{La:Lif}.}\ solutions (\ref{WaveFunctionN})
do satisfy the time-dependent Schr\"{o}dinger equation (\ref{Schroudinger}).
Indeed, they are written as the linear superposition (\ref%
{ExpansionFinalIndependent})--(\ref{ExpansionCoeffs}) of standard solutions.

\section{Nonclassical Harmonic States of Motion and Photon Statistics}

A fundamental manifestation of the interaction between an atom and a field
mode at resonance in an ideal cavity is the Rabi oscillations \cite%
{HarocheRaimond06}. The first observation of the nonclassical radiation
field of a micromaser is reported in \cite{Rempeetal90} (the statistical and
discrete nature of the photon field leads to collapse and revivals in the
Rabi nutation \cite{RempeWaltherKlein87}). Implementation of light for
purposes of quantum information relies on the ability to synthesize,
manipulate, and characterize various quantum states of the electromagnetic
field. A review \cite{LvRay09} covers the latest developments in
quantum-state tomography of optical fields and photons (see also the
references therein).

Various classes of motional states in ion traps are discussed, for example,
in \cite{LeibfriedetalWineland03}. Our expansion formula (\ref%
{ExpansionCoeffs}) is consistent with statistics for the coherent, squeezed,
and Fock states observed in Refs.~\cite{Breit:Schill:Mlyn97} and \cite%
{Meekhofetal96} for ions and photons in a box (see also \cite{Dod:Man:Man94}%
, \cite{Kimetal89} and \cite{LeibfriedetalWineland03}). A method to measure
the quantum state of a harmonic oscillator through instantaneous
probe-system interaction, preventing decoherence from disturbing the
measurement, is proposed in \cite{Santosetal07}.

\subsection{Coherent States}

In breakthrough experiments of the NIST group on engineering ionic states of
motion, the coherent states of a single $^{9}Be^{+}$ ion confined in a Paul
trap were produced from the ground state by a spatially uniform classical
driving field and by \textquotedblleft moving standing
wave\textquotedblright\ (see \cite{LeibfriedetalWineland03}, \cite%
{Meekhofetal96} and the references therein for details). For the data
presented in \cite{Meekhofetal96}, the authors used the first method. The
Poissonian distribution with the fitted mean quantum number $\overline{n}%
=3.1\pm 0.1$ was identified from Fourier analysis of Rabi oscillations. In
our notation, $\alpha _{0}=0,$ $\beta _{0}=1,$ and $\overline{n}=\left(
\delta _{0}^{2}+\varepsilon _{0}^{2}\right) /2.$

Time evolution of the coherent state of cold Cs atoms was measured in \cite%
{Morinagaetal99}. For experimentally observed coherent photon states \cite%
{GlauberCollect}; see, for example, \cite{Breit:Schill:Mlyn97} and \cite%
{LvovskyBabichev02}.

\subsection{Squeezed Vacuum and Fock States}

The minimum-uncertainty squeezed state with $\gamma _{0}=\delta
_{0}=\varepsilon _{0}=\kappa _{0}=0$ is called the squeezed vacuum (see \cite%
{Dod:Man:Man94}, \cite{Kimetal89}, \cite{KlauderSudarshan}, and \cite%
{LeibfriedetalWineland03} when $\alpha _{0}=0).$ Expansion (\ref%
{ExpansionFinalIndependent}) simplifies to%
\begin{eqnarray}
&&\psi _{0}\left( x,t\right) =e^{i\left( \alpha \left( t\right) x^{2}+\gamma
\left( t\right) \right) }\sqrt{\frac{\beta \left( t\right) }{\sqrt{\pi }}}\
e^{-\beta ^{2}\left( t\right) x^{2}/2}  \label{SqueezedVacuum} \\
&&\ =\sqrt{\beta _{0}}\sum_{p=0}^{\infty }\frac{\sqrt{\left( 2p\right) !}}{%
2^{p}p!}\frac{\left( \dfrac{1-\beta _{0}^{2}}{2}+i\alpha _{0}\right) ^{p}}{%
\left( \dfrac{1+\beta _{0}^{2}}{2}-i\alpha _{0}\right) ^{p+1/2}}\
e^{-i\left( 2p+1/2\right) t}\ \Psi _{2p}\left( x\right) .  \notag
\end{eqnarray}%
The probability distribution is restricted to the even states and given by%
\begin{equation}
P_{m=2p}=\frac{\left( 2p\right) !}{\left( \sigma _{p}+\sigma _{x}+1\right)
^{1/2}2^{2p-1/2}\left( p!\right) ^{2}}\left( \frac{\sigma _{p}+\sigma _{x}-1%
}{\sigma _{p}+\sigma _{x}+1}\right) ^{p}  \label{Pascal0}
\end{equation}%
in terms of the variances (\ref{InvariantsSigma}). This is a special case of
the negative binomial, or Pascal, distribution.

A vacuum squeezed state of ionic motion was created in the NIST group
experiments \cite{Meekhofetal96} by a parametric drive at $2\nu $ (see also 
\cite{HeinzenWineland90}, \cite{LeibfriedetalWineland03} and the references
therein). The data were fitted to the vacuum state distribution (\ref%
{Pascal0}) with $\sigma _{p}+\sigma _{x}=40\pm 10$ and $\alpha _{0}=0$
(corresponding to a noise level $16~$dB below the zero-point variance in the
squeezed quadrature component; see \cite{LeibfriedetalWineland03} and \cite%
{Meekhofetal96} for more experimental details).

A vacuum squeezed state of motion of neutral Cs atoms was also generated in 
\cite{Morinagaetal99}. Here, the cold atom sample containes about $10^{5}$
atoms. Therefore a single image provides the full velocity distribution of
the quantum state and the squeezing can be readily visualized --- a set of
images gives the state's time evolution (see \cite{Morinagaetal99} and the
references therein for more details).

In a similar fashion, for the squeezed Fock state with $n=1$ and $\gamma
_{0}=\delta _{0}=\varepsilon _{0}=\kappa _{0}=0,$ expansion (\ref%
{ExpansionFinalIndependent}) simplifies to%
\begin{eqnarray}
&&\psi _{1}\left( x,t\right) =\sqrt{\frac{2\beta \left( t\right) }{\sqrt{\pi 
}}}e^{i\left( \alpha \left( t\right) x^{2}+3\gamma \left( t\right) \right)
}\beta \left( t\right) xe^{-\beta ^{2}\left( t\right) x^{2}/2}
\label{SqueezedFock} \\
&&\ =\frac{\beta _{0}^{3/2}}{\sqrt{\pi }}\sum_{p=0}^{\infty }\frac{%
2^{p+1}\Gamma \left( p+3/2\right) }{\sqrt{\left( 2p+1\right) !}}\frac{\left( 
\dfrac{1-\beta _{0}^{2}}{2}+i\alpha _{0}\right) ^{p}}{\left( \dfrac{1+\beta
_{0}^{2}}{2}-i\alpha _{0}\right) ^{p+3/2}}\ e^{-i\left( 2p+3/2\right) t}\
\Psi _{2p+1}\left( x\right) .  \notag
\end{eqnarray}%
The corresponding Pascal distribution for the odd states is given by%
\begin{equation}
P_{m=2p+1}=\frac{2^{3/2}\left( 3/2\right) _{p}}{\left( \sigma _{p}+\sigma
_{x}+1\right) ^{3/2}p!}\left( \frac{\sigma _{p}+\sigma _{x}-1}{\sigma
_{p}+\sigma _{x}+1}\right) ^{p},  \label{Pascal1}
\end{equation}%
where $\left( 3/2\right) _{0}=1$ and $\left( 3/2\right) _{p}=\left(
3/2\right) \left( 5/2\right) \cdot \cdot \cdot \left( 1/2+p\right) .$ These
squeezed Fock states were generated in \cite{Bouchouleetal99} and their
dynamics was analyzed in \cite{Morinagaetal99}. When $\varepsilon _{0}\neq
0, $ displaced Fock states of the electromagnetic field, have been
synthesized in \cite{LvovskyBabichev02} (see also the references therein).

Moreover, even/odd oscillations in the photon number distribution of the
\textquotedblleft squeezed vacuum\textquotedblright\ state, which are
consequence of pair-wise generation of photon, were observed in \cite%
{Breit:Schill:Mlyn97}, \cite{SchillerBreitenbachetal96}. For an ideal
minimum-uncertainty squeezed state zero probabilities for odd $n$ are
expected, since the Hamiltonian describing the parametric process occurring
inside the nonlinear crystal is quadratic in the creation and annihilation
operators \cite{Dod:Man:Man94}, \cite{Schleich01}. However, the
probabilities for odd photon numbers are nonzero because the squeezed state
detected there is a mixed state having undergone losses inside the resonator
and during the detection process which cause the distribution to smear out
(see \cite{Dod:Man:Man94} and \cite{SchillerBreitenbachetal96} for more
details). The corresponding Pascal distributions (\ref{Pascal0}) and (\ref%
{Pascal1}) have different parameter values for even and odd states, which is
consistent with the result of these experiments. Further details will be
discussed elsewhere.

\subsection{Superposition of Fock States}

Generation of a coherent superposition of the ground state and the first
excited Fock states of motion of cold Cs atoms in the harmonic microtraps,
namely,%
\begin{equation}
\psi \left( x,t\right) =c_{0}e^{-it/2}\ \Psi _{0}\left( x\right)
+c_{1}e^{-3it/2}\ \Psi _{1}\left( x\right)  \label{Lin2Ionic}
\end{equation}%
where $c_{0}=2^{-1/2}$ and $c_{1}=2^{-1/2}e^{i\phi },$ was reported in \cite%
{Morinagaetal99} and the corresponding time evolution had been
experimentally observed. This nonclassical evolution contrasts with that of
a coherent state which oscillates as a classical particle without
deformation (see \cite{Morinagaetal99} for more details).

\subsection{Superfocusing of Particle Beams}

An effect of proton beam focusing in a thin monocrystal film was predicted
in Refs.~\cite{Demkov09}, \cite{DemkovMeyer04}. A highly collimated beam of
protons ($\approx 1~$MeV) entering the channel of a monocrystal film forms
at a certain depth an extremely sharp ($<0.005~$nm) and relatively long
(some monolayers of the crystal) focusing area where the increase of the
flux density can reach up to thousand times. We shall refer to this effect
as superfocusing (or Demkov's microscope). The mean effective potential of
the channel can be calculated and the deflection of the fast particle within
the channel can be found. In many cases the potential of the central part of
the averaged channel is cylindrically symmetric and harmonic to a good
approximation which can create isochronic oscillations of the ions in the
plane normal to the direction of the channel. The radius of this focus can,
in principle, be as small as $10^{-2}~$nm.

According to Demkov's theoretical model \cite{Demkov09}, the channel average
potential is independent of the channel direction $z$ and can be
approximated by $\left( x^{2}+y^{2}\right) /2$ for the transverse direction.
The $z$ motion along the channel is treated classically which allows one to
replace $z$ by the time $t$ setting the velocity equal to unity. By
separation of variables, the normalized $2D$ time-dependent Schr\"{o}dinger
equation,%
\begin{equation}
2i\psi _{t}+\psi _{xx}+\psi _{yy}=\left( x^{2}+y^{2}\right) \psi ,
\label{Demkov2D}
\end{equation}%
has the following orthonormal solution:%
\begin{eqnarray}
&&\psi \left( x,y,t\right) =\frac{e^{-i\arctan \left( \beta _{0}^{2}\tan
t\right) }}{\sqrt{\pi }}\left( \beta _{0}^{2}\sin ^{2}t+\beta _{0}^{-2}\cos
^{2}t\right) ^{-1/2}  \label{DemkovSolution} \\
&&\qquad \qquad \quad \times \exp \left( i\frac{\left( \beta _{0}^{2}-\beta
_{0}^{-2}\right) \left( x^{2}+y^{2}\right) \sin 2t}{4\left( \beta
_{0}^{2}\sin ^{2}t+\beta _{0}^{-2}\cos ^{2}t\right) }\right)  \notag \\
&&\qquad \qquad \quad \quad \times \exp \left( i\frac{\delta _{0}\left(
2x-\delta _{0}\sin t\right) \cos t}{2\beta _{0}^{2}\left( \beta _{0}^{2}\sin
^{2}t+\beta _{0}^{-2}\cos ^{2}t\right) }\right)  \notag \\
&&\qquad \qquad \qquad \quad \times \exp \left( -\frac{\left( x-\delta
_{0}\sin t\right) ^{2}+y^{2}}{2\left( \beta _{0}^{2}\sin ^{2}t+\beta
_{0}^{-2}\cos ^{2}t\right) }\right)  \notag
\end{eqnarray}%
(the minimum-uncertainty squeezed state). Then%
\begin{equation}
\left\vert \psi \left( \boldsymbol{r}_{\perp },t\right) \right\vert
^{2}=\left( \pi \left( \beta _{0}^{2}\sin ^{2}t+\beta _{0}^{-2}\cos
^{2}t\right) \right) ^{-1}\exp \left( -\frac{\left( x-\delta _{0}\sin
t\right) ^{2}+y^{2}}{\beta _{0}^{2}\sin ^{2}t+\beta _{0}^{-2}\cos ^{2}t}%
\right)  \label{DemkovNorm}
\end{equation}%
with $\delta _{0}=-p_{x}$ and $\beta _{0}=R_{\text{min}},$ $\beta
_{0}^{-1}=R_{\text{s}}.$ Here, $R_{\text{min}}R_{\text{s}}=1$ in the units
of original papers \cite{Demkov09}, \cite{DemkovMeyer04}.

Among other things, Demkov has predicted that the counter beams may raise an
yield of nuclear reactions by orders of magnitude. He also proposed an idea
of grouping of beams under action of longitudinal sawtooth fluctuations of
accelerating potential. The validity of his $2D$ harmonic channel model was
confirmed by Monte Carlo computer experiments \cite{Demkov09}, \cite%
{DemkovMeyer04}. (For analogous lens effects in paraxial optics; see \cite%
{AbramVolUFN}, \cite{Agrawaletal74}, \cite{MahSusParOptics},\ \cite%
{VinRudSuxBook79} and the references therein.)

\section{An Application to Cavity QED and Quantum Optics}

Foundations of quantum electrodynamics and quantum optics are presented in
many excellent books and articles \cite{Akh:Ber}, \cite{Ber:Lif:Pit}, \cite%
{Bia:Bia75}, \cite{Bo:Shi}, \cite{DiracQM}, \cite{Dod:Klim:Nik93}, \cite%
{DutraQED}, \cite{FeynmanFundamental}, \cite{FeynmanQED}, \cite{Fey:Hib}, 
\cite{FermiRad}, \cite{Fermi}, \cite{Fock32-2}, \cite{Fock34-3}, \cite%
{GlauberCollect}, \cite{Glauber91}, \cite{HeisenbergQM}, \cite{It:Zu}, \cite%
{JaynesCummings63}, \cite{Klauder12}, \cite{KlauderSudarshan}, \cite%
{Louisell73}, \cite{Merz}, \cite{Rowe77}, \cite{Schiff}, \cite%
{Sch:Plun:Soff98}, \cite{Slater1950}, \cite{Walls83}, \cite{WallsMilburn08}, 
\cite{Wein}. Here, we suggest a modification of the radiation field
operators in a perfect cavity in order to incorporate the Schr\"{o}dinger
symmetry group into the second quantization. Our approach gives a natural
description of squeezed photons that can be created as a result of
parametric amplification of quantum fluctuations in the dynamic Casimir
effect \cite{Lahetal11}, \cite{Wilsonetal11} and are registered in quantum
optics \cite{Breit:Schill:Mlyn97}, \cite{LvRay09}, \cite{Ourjoumtsevetal11}.

\subsection{Radiation Field Quantization in a Perfect Cavity}

In the formalism of second quantization, one expands electromagnetic fields
in terms of resonant modes of the particular cavity under consideration \cite%
{DutraQED}, \cite{Glauber91}, \cite{JaynesCummings63}, \cite{Schleich01}, 
\cite{Slater1950}. The cavity is represented by a volume $V,$ bounded by a
closed surface. Let $\boldsymbol{E}_{\lambda }\left( \boldsymbol{r}\right) ,$
$k_{\lambda }^{2}=\omega _{\lambda }^{2}/c^{2}$ be the eigenfunctions and
the eigenvalues of the corresponding boundary-value problem:%
\begin{eqnarray}
\boldsymbol{\nabla \times \nabla \times E}-k^{2}\boldsymbol{E}\boldsymbol{=0}
&&\qquad \text{in }V  \label{EigenValueProblemCavity} \\
\boldsymbol{n\times E}\boldsymbol{=0} &&\text{\qquad on }S,  \notag
\end{eqnarray}%
where $\boldsymbol{n}$ is a unit normal vector to $S.$ The vector functions $%
\boldsymbol{H}_{\lambda }\left( \boldsymbol{r}\right) $ are related to $%
\boldsymbol{E}_{\lambda }\left( \boldsymbol{r}\right) $ by%
\begin{equation}
\boldsymbol{\nabla \times E}_{\lambda }=k_{\lambda }\boldsymbol{H}_{\lambda
},\qquad \boldsymbol{\nabla \times H}_{\lambda }=k_{\lambda }\boldsymbol{E}%
_{\lambda }.  \label{VectorsEH}
\end{equation}%
The eigenfunctions are orthonormal in $V:$%
\begin{equation}
\int_{V}\boldsymbol{E}_{\lambda }\cdot \boldsymbol{E}_{\mu }\ dV=\delta
_{\lambda \mu },\qquad \int_{V}\boldsymbol{H}_{\lambda }\cdot \boldsymbol{H}%
_{\mu }\ dV=\delta _{\lambda \mu }.  \label{OrthogonaltyEH}
\end{equation}%
The electric and magnetic fields are expanded in the following forms%
\begin{eqnarray}
&&\boldsymbol{E}\left( \boldsymbol{r},t\right) =-\sqrt{4\pi }\sum_{\lambda
}p_{\lambda }\left( t\right) \boldsymbol{E}_{\lambda }\left( \boldsymbol{r}%
\right) ,  \label{CavityExpansionsEH} \\
&&\boldsymbol{H}\left( \boldsymbol{r},t\right) =\sqrt{4\pi }\sum_{\lambda
}\omega _{\lambda }q_{\lambda }\left( t\right) \boldsymbol{H}_{\lambda
}\left( \boldsymbol{r}\right) .  \notag
\end{eqnarray}%
The total energy is given by%
\begin{equation}
\mathcal{H}=\int \frac{\boldsymbol{H}^{2}+\boldsymbol{E}^{2}}{8\pi }\ dV=%
\frac{1}{2}\sum_{\lambda }\left( p_{\lambda }^{2}+\omega _{\lambda
}^{2}q_{\lambda }^{2}\right)   \label{EMHamiltonian}
\end{equation}%
and the Maxwell equations,%
\begin{equation}
\boldsymbol{\nabla \times E}=-\frac{1}{c}\frac{\partial \boldsymbol{H}}{%
\partial t},\qquad \boldsymbol{\nabla \times H}=\frac{1}{c}\frac{\partial 
\boldsymbol{E}}{\partial t},  \label{MaxwellTwo}
\end{equation}%
are equivalent to the canonical Hamiltonian equations,%
\begin{equation}
\frac{dq_{\lambda }}{dt}=\frac{\partial \mathcal{H}}{\partial p_{\lambda }}%
=p_{\lambda },\qquad \frac{dp_{\lambda }}{dt}=-\frac{\partial \mathcal{H}}{%
\partial q_{\lambda }}=-\omega _{\lambda }^{2}q_{\lambda },
\label{HamiltonianEquationsPQ}
\end{equation}%
respectively.

In the second quantization, one replaces canonically conjugate coordinates
and momenta by time-dependent operators $q_{\lambda }\left( t\right) $ and $%
p_{\lambda }\left( t\right) $ that satisfy the commutation rules%
\begin{equation}
\left[ q_{\lambda }\left( t\right) ,\ q_{\mu }\left( t\right) \right] =\left[
p_{\lambda }\left( t\right) ,\ p_{\mu }\left( t\right) \right] =0,\qquad %
\left[ q_{\lambda }\left( t\right) ,\ p_{\mu }\left( t\right) \right]
=i\hbar \delta _{\lambda \mu }.  \label{CommutatorsPQ}
\end{equation}%
The time-evolution is determined by the Heisenberg equations of motion \cite%
{HeisenbergQM}:%
\begin{equation}
\frac{d}{dt}p_{\lambda }\left( t\right) =\frac{i}{\hbar }\left[ p_{\lambda
}\left( t\right) ,\ \mathcal{H}\right] ,\qquad \frac{d}{dt}q_{\lambda
}\left( t\right) =\frac{i}{\hbar }\left[ q_{\lambda }\left( t\right) ,\ 
\mathcal{H}\right] ,  \label{HeisenbergEquationsPQ}
\end{equation}%
with appropriate initial conditions.\footnote{%
The standard form of Heisenberg's equations can be obtained by the time
reversal $t\rightarrow -t$ (with $\alpha _{0}\rightarrow -\alpha _{0},$ $%
\gamma _{0}\rightarrow -\gamma _{0},$ $\delta _{0}\rightarrow -\delta _{0},$
and $\kappa _{0}\rightarrow -\kappa _{0}).$} (From now on, we consider a
single photon cavity mode, say $\upsilon ,$ with frequency $\omega
_{\upsilon }=1$ and use the units $c=\hbar =1.)$

\subsection{Nonstandard Solutions of Heisenberg's Equations}

Explicit solution of equations (\ref{HeisenbergEquationsPQ}) for squeezed
states can be found as follows%
\begin{equation}
p\left( t\right) =\frac{\widehat{b}\left( t\right) -\widehat{b}^{\dagger
}\left( t\right) }{i\sqrt{2}},\qquad q\left( t\right) =\frac{\widehat{b}%
\left( t\right) +\widehat{b}^{\dagger }\left( t\right) }{\sqrt{2}}.
\label{pqQED}
\end{equation}%
The time-dependent annihilation $\widehat{b}\left( t\right) $ and creation $%
\widehat{b}^{\dagger }\left( t\right) $ operators are given by \cite%
{Kr:Sus12}%
\begin{eqnarray}
&&\widehat{b}\left( t\right) =\frac{e^{-2i\gamma }}{\sqrt{2}}\left( \beta
x+\varepsilon +i\frac{p-2\alpha x-\delta }{\beta }\right) ,
\label{aacross(t)QED} \\
&&\widehat{b}^{\dagger }\left( t\right) =\frac{e^{2i\gamma }}{\sqrt{2}}%
\left( \beta x+\varepsilon -i\frac{p-2\alpha x-\delta }{\beta }\right) 
\notag
\end{eqnarray}%
\noindent in terms of the solutions (\ref{hhA})--(\ref{hhK}) of Ermakov-type
system. The time-independent operators $x$ and $p$ obey the canonical
commutation rule $\left[ x,p\right] =i$ in an abstract Hilbert space. At all
times,%
\begin{equation}
\widehat{b}\left( t\right) \widehat{b}^{\dagger }\left( t\right) -\widehat{b}%
^{\dagger }\left( t\right) \widehat{b}\left( t\right) =1.
\label{commutatora(t)across(t)QED}
\end{equation}%
By back substitution, operators $\widehat{b}\left( t\right) $ and $\widehat{b%
}^{\dagger }\left( t\right) $ are solutions of the Heisenberg equation:%
\begin{equation}
\frac{d}{dt}\widehat{b}\left( t\right) =i\left[ \ \widehat{b}\left( t\right)
,\ H\right] ,\qquad \frac{d}{dt}\widehat{b}^{\dagger }\left( t\right) =i%
\left[ \ \widehat{b}^{\dagger }\left( t\right) ,\ H\right] ,
\label{HeisenberEquations}
\end{equation}%
with the standard Hamiltonian%
\begin{equation}
H=\frac{1}{2}\left( p^{2}+x^{2}\right)  \label{Hamiltonian}
\end{equation}%
subject to the following initial conditions%
\begin{eqnarray}
&&\widehat{b}\left( 0\right) =\frac{e^{-2i\gamma _{0}}}{\sqrt{2}}\left(
\beta _{0}x+\varepsilon _{0}+i\frac{p-2\alpha _{0}x-\delta _{0}}{\beta _{0}}%
\right) ,  \label{HeisenbergInitialData} \\
&&\widehat{b}^{\dagger }\left( 0\right) =\frac{e^{2i\gamma _{0}}}{\sqrt{2}}%
\left( \beta _{0}x+\varepsilon _{0}-i\frac{p-2\alpha _{0}x-\delta _{0}}{%
\beta _{0}}\right) .  \notag
\end{eqnarray}%
One may say that the transformation (\ref{aacross(t)QED}) allow us to
incorporate the Schr\"{o}dinger group of harmonic oscillator, originally
found in coordinate representation \cite{Niederer73}, into a more abstract
Heisenberg picture (the classical case occurs when $\beta _{0}=1$ and $%
\alpha _{0}=\gamma _{0}=\delta _{0}=\varepsilon _{0}=\kappa _{0}=0).$

\subsection{Dynamic Fock Space for a Single Mode}

The time-dependent quadratic invariant,%
\begin{eqnarray}
\widehat{E}\left( t\right) &=&\frac{1}{2}\left[ \frac{\left( p-2\alpha
x-\delta \right) ^{2}}{\beta ^{2}}+\left( \beta x+\varepsilon \right) ^{2}%
\right]  \label{QIQED} \\
&=&\frac{1}{2}\left[ \widehat{b}\left( t\right) \widehat{b}^{\dagger }\left(
t\right) +\widehat{b}^{\dagger }\left( t\right) \widehat{b}\left( t\right) %
\right] ,\qquad \frac{d}{dt}\langle \widehat{E}\left( t\right) \rangle =0 
\notag
\end{eqnarray}%
with%
\begin{equation}
\frac{\partial \widehat{E}}{\partial t}+i^{-1}\left[ \widehat{E},H\right]
=0,\qquad H=\frac{1}{2}\left( p^{2}+x^{2}\right) ,  \label{QDQED}
\end{equation}%
extends the standard Hamiltonian/Number operator $H$ for any given real
values of parameters/integrals of motion in our description of the squeezed
photon state. The oscillator-type spectrum,%
\begin{equation}
\widehat{E}\left( t\right) \left\vert \psi _{n}\left( t\right) \right\rangle
=\left( n+\frac{1}{2}\right) \left\vert \psi _{n}\left( t\right)
\right\rangle ,  \label{EigenValueProblemQED}
\end{equation}%
can be obtained by using the modified creation and annihilation operators 
\cite{Akh:Ber}:%
\begin{eqnarray}
&&\widehat{b}\left( t\right) \left\vert \psi _{n}\left( t\right)
\right\rangle =\sqrt{n}\ \left\vert \psi _{n-1}\left( t\right) \right\rangle
,  \label{annandcratoperactionsQED} \\
&&\widehat{b}^{\dagger }\left( t\right) \left\vert \psi _{n}\left( t\right)
\right\rangle =\sqrt{n+1}\ \left\vert \psi _{n+1}\left( t\right)
\right\rangle .  \notag
\end{eqnarray}%
With a proper choice of the global phase, the latter eigenstates of
dynamical invariant satisfy the time-dependent Schr\"{o}dinger equation in
an abstract Hilbert space \cite{Fock28-2}, \cite{Kr:Sus12}.

For the \textquotedblleft minimum-uncertainty squeezed
states\textquotedblright , one gets 
\begin{equation}
\widehat{b}\left( t\right) \left\vert \psi _{0}\left( t\right) \right\rangle
=0  \label{VacuumQED}
\end{equation}%
with%
\begin{equation}
\langle \psi _{0}\left( t\right) \left\vert H\right\vert \psi _{0}\left(
t\right) \rangle =\dfrac{1+4\alpha _{0}^{2}+\beta _{0}^{4}}{4\beta _{0}^{2}}+%
\frac{\left( 2\alpha _{0}\varepsilon _{0}-\beta _{0}\delta _{0}\right)
^{2}+\varepsilon _{0}^{2}}{2\beta _{0}^{2}}\geq \frac{1}{2}
\label{SqueezedEnergy}
\end{equation}%
in the Schr\"{o}dinger picture. The generalized coherent (or TCS's) states
are given by%
\begin{equation}
\widehat{b}\left( t\right) \left\vert \psi \left( t\right) \right\rangle
=\zeta \left\vert \psi \left( t\right) \right\rangle  \label{TCSQED}
\end{equation}%
for an arbitrary complex $\zeta \neq 0.$

\subsection{Expectation Values and Variances for Field Oscillators}

The noncommuting electric $\boldsymbol{E}\left( \boldsymbol{r},t\right) $
and magnetic $\boldsymbol{H}\left( \boldsymbol{r},t\right) $ field operators
are given by equations (\ref{CavityExpansionsEH}) and (\ref{pqQED})--(\ref%
{aacross(t)QED}) for a squeezed photon in the Heisenberg picture, which
provides a more direct analogy between quantum and classical physics \cite%
{HarocheRaimond06}. The electromagnetic radiation mode in a cavity resonator
is analogous to a harmonic oscillator \cite{Henry:Glotzer88}. In the Schr%
\"{o}dinger picture, all previous results on the minimum-uncertainty
squeezed states can be reproduced for the field oscillators in an operator
QED-style. For a single mode with $\omega _{\upsilon }=1,$%
\begin{eqnarray}
&&\left\langle \boldsymbol{E}\left( \boldsymbol{r},t\right) \right\rangle =-%
\sqrt{4\pi }\boldsymbol{E}_{\upsilon }\left( \boldsymbol{r}\right)
\left\langle \psi _{n}\left( t\right) \left\vert p\right\vert \psi
_{n}\left( t\right) \right\rangle ,  \label{EHradiationOperators} \\
&&\left\langle \boldsymbol{H}\left( \boldsymbol{r},t\right) \right\rangle =%
\sqrt{4\pi }\boldsymbol{H}_{\upsilon }\left( \boldsymbol{r}\right)
\left\langle \psi _{n}\left( t\right) \left\vert x\right\vert \psi
_{n}\left( t\right) \right\rangle ,  \notag
\end{eqnarray}%
where equations (\ref{<x>})--(\ref{<p>}) hold. The corresponding variances
are given (up to a normalization) by equations (A.4)--(A.5) of Ref.~\cite%
{LopSusVegaHarm}.

The minimum-uncertainty squeezed states are identified in quantum optics 
\cite{Dodonov02}, \cite{Henry:Glotzer88}, \cite{Guerr:Lop:Ald:Coss11}, \cite%
{JainetalLv10}, \cite{Kimetal89}, \cite{LeibfriedetalWineland03}, \cite%
{Shchukinetal09}, \cite{Slusheretal85}, \cite{RiesetalLv03}, \cite{Rowe77}, 
\cite{Yuen76} and in state tomography \cite{ChernegaManko11}, \cite%
{Eichleretal11}, \cite{LeonardPaul95}, \cite{LvRay09}. They are also
important in the dynamical Casimir effect \cite{Dodonov10}, \cite%
{DodKlimMan90}, \cite{Dod:Klim:Nik93}, \cite{DutraQED}, \cite%
{Fu:Mat:Hat:Kur:Zeil}, \cite{Kr:Sus12}, \cite{Lahetal11}, \cite%
{Man'koCasimir}, \cite{Wilsonetal11}, and \cite{You:Nori11}, where the
photon squeezing occurs as a result of a \textquotedblleft parametric
excitation\textquotedblright\ of vacuum oscillations.

\section{An Important Variance}

The Hamiltonian $H=\left( p^{2}+x^{2}\right) /2$ can be rewritten in terms
of the creation and annihilation operators (\ref{a(t)}) as follows: 
\begin{eqnarray}
H &=&\left( \frac{4\alpha ^{2}-\beta ^{4}+1}{4\beta ^{2}}-i\alpha \right) 
\widehat{a}^{2}\left( t\right) +\left( \frac{4\alpha ^{2}-\beta ^{4}+1}{%
4\beta ^{2}}+i\alpha \right) \left. \widehat{a}^{\dagger }\left( t\right)
\right. ^{2}  \label{HAB} \\
&&+\frac{4\alpha ^{2}+\beta ^{4}+1}{4\beta ^{2}}\left[ \widehat{a}\left(
t\right) \widehat{a}^{\dagger }\left( t\right) +\widehat{a}^{\dagger }\left(
t\right) \widehat{a}\left( t\right) \right]  \notag \\
&&+\sqrt{2}\left[ \frac{\alpha }{\beta }\left( \delta -\frac{2\alpha
\varepsilon }{\beta }\right) -\frac{\varepsilon }{2\beta ^{2}}-\frac{i\beta 
}{2}\left( \delta -\frac{2\alpha \varepsilon }{\beta }\right) \right] 
\widehat{a}\left( t\right)  \notag \\
&&+\sqrt{2}\left[ \frac{\alpha }{\beta }\left( \delta -\frac{2\alpha
\varepsilon }{\beta }\right) -\frac{\varepsilon }{2\beta ^{2}}+\frac{i\beta 
}{2}\left( \delta -\frac{2\alpha \varepsilon }{\beta }\right) \right] 
\widehat{a}^{\dagger }\left( t\right)  \notag \\
&&+\frac{1}{2}\left( \delta -\frac{2\alpha \varepsilon }{\beta }\right) ^{2}+%
\frac{\varepsilon ^{2}}{2\beta ^{2}}  \notag
\end{eqnarray}%
and by definition:%
\begin{equation}
\text{Var\ }H=\langle \left( H-\langle H\rangle \right) ^{2}\rangle =\langle
H^{2}\rangle -\langle H\rangle ^{2}.  \label{VarH}
\end{equation}%
Then a direct \textsl{Mathematica} calculation results in%
\begin{eqnarray}
&&\text{Var\ }H=\frac{\left( 4\alpha _{0}^{2}+\left( \beta _{0}^{2}+1\right)
^{2}\right) \left( 4\alpha _{0}^{2}+\left( \beta _{0}^{2}-1\right)
^{2}\right) }{8\beta _{0}^{4}}\left[ \left( n+\frac{1}{2}\right) ^{2}+\frac{3%
}{4}\right]  \label{VarHFinal} \\
&&+\left[ \frac{\left( 4\alpha _{0}^{2}+\beta _{0}^{4}+1\right) \left(
\left( 2\alpha _{0}\varepsilon _{0}-\beta _{0}\delta _{0}\right)
^{2}+\varepsilon _{0}^{2}\right) }{\beta _{0}^{4}}-\left( \varepsilon
_{0}^{2}+\frac{\delta _{0}^{2}}{\beta _{0}^{2}}\right) \right] \left( n+%
\frac{1}{2}\right)  \notag
\end{eqnarray}%
for the wavefunctions (\ref{WaveFunctionN}) in terms of the invariants (\ref%
{Invs1})--(\ref{Invs2}). (These calculations can be performed in pure
operator form with the help of standard relations (1.15) of Ref.~\cite%
{LopSusVegaHarm}; see also (\ref{annandcratoperactionsQED}).) In terms of
the variances,%
\begin{eqnarray}
\text{Var\ }H &=&\frac{1}{2}\left[ \left( \sigma _{p}+\sigma _{x}\right)
^{2}-1\right] \left[ \left( n+\frac{1}{2}\right) ^{2}+\frac{3}{4}\right]
\label{VarHInvariants} \\
&&+2\left[ \sigma _{p}\langle p\rangle ^{2}+2\sigma _{px}\langle p\rangle
\langle x\rangle +\sigma _{x}\langle x\rangle ^{2}\right] \left( n+\frac{1}{2%
}\right) ,  \notag
\end{eqnarray}%
where $\sigma _{p},$ $\sigma _{x},$ and $\sigma _{px}$ are given by (\ref%
{Variences}). When $n=0,$ this formula is consistent with the variance of
the number operator derived for a generic Gaussian Wigner function in Ref.~%
\cite{Dod:Man:Man94}.

A similar expression holds for the TCS states. Computational details are
left to the reader.

\section{A Conclusion}

In this paper, we review some properties of the nonclassical states of
harmonic motion which were originally found in \cite{Marhic78} (in
coordinate representation) and have been rediscovered recently in \cite%
{LopSusVegaHarm}. They are useful in applications to cavity QED, quantum
optics, and in channeling scattering \cite{Demkov09}. In particular, the
minimum-uncertainty squeezed states are studied in detail. Expansions in the
Fock states are established and their relations with experimentally observed
photon statistics are briefly discussed. In the method of second
quantization, a modification of the radiation field operators for squeezed
photons in a perfect cavity is suggested with the help of a nonstandard
solution of Heisenberg's equation of motion. These results may be of
interest to everyone who studies introductory quantum mechanics and quantum
optics. \medskip

\noindent \textbf{Acknowledgments.\/} We would like to thank Viktor Dodonov,
Mark Faifman, Geza Giedke, John R.~Klauder, Alex Mahalov, Vladimir I.
Man'ko, and Andreas Ruffing for valuable discussions. The authors are
grateful to the referees for valuable suggestions. This research is
supported in part by the National Science Foundation--Enhancing the
Mathematical Sciences Workforce in the 21st Century (EMSW21), award \#
0838705; the Alfred P. Sloan Foundation--Sloan National Pipeline Program in
the Mathematical and Statistical Sciences, award \# LTR 05/19/09.

\appendix

\section{A Complex Parametrization of the Schr\"{o}dinger Group}

The Ansatz 
\begin{equation}
\psi \left( x,t\right) =\sqrt{\beta \left( t\right) }e^{iS\left( x,t\right)
}\ \chi \left( \xi ,\tau \right) ,\qquad S=\alpha \left( t\right)
x^{2}+\delta \left( t\right) x+\kappa \left( t\right) ,
\label{SchroedingerOscillator}
\end{equation}%
where relations (\ref{hhA})--(\ref{hhK}) hold, transforms the time-dependent
Schr\"{o}dinger equation (\ref{Schroudinger}) into itself:%
\begin{equation}
2i\psi _{t}+\psi _{xx}-x^{2}\psi =e^{iS}\beta ^{5/2}\left( 2i\chi _{\tau
}+\chi _{\xi \xi }-\xi ^{2}\chi \right) =0  \label{SchroedingerGroup}
\end{equation}%
with respect to the new variables $\xi =\beta \left( t\right) x+\varepsilon
\left( t\right) $ and $\tau =-\gamma \left( t\right) .$ This transformation
is known as the Schr\"{o}dinger group for linear harmonic oscillator \cite%
{Niederer73}.

Let us introduce the following complex-valued function:%
\begin{equation}
z=c_{1}e^{it}+c_{2}e^{-it},\qquad z^{\prime \prime }+z=0,  \label{zeta}
\end{equation}%
where by definition%
\begin{eqnarray}
&&c_{1}=\left( 1+\beta _{0}^{2}\right) /2-i\alpha _{0},\qquad c_{2}=\left(
1-\beta _{0}^{2}\right) /2+i\alpha _{0}  \label{zeta12} \\
&&\quad \left( c_{1}+c_{2}=1,\qquad \quad \left\vert c_{1}\right\vert
^{2}-\left\vert c_{2}\right\vert ^{2}=\beta _{0}^{2}\right) ,  \notag
\end{eqnarray}%
and%
\begin{equation}
c_{3}=\frac{\delta _{0}}{\beta _{0}}-i\varepsilon _{0}.  \label{zeta3}
\end{equation}%
Then equations (\ref{hhA})--(\ref{hhK}) can be rewritten in a compact form
in terms of our complex parameters $c_{1},$ $c_{2},$ and $c_{3}.$ Indeed,
with the help of identities (\ref{Id1})--(\ref{Id4}), one gets%
\begin{equation}
\left\vert z\right\vert =\left( \left\vert c_{1}\right\vert
^{2}+c_{1}c_{2}^{\ast }e^{2it}+c_{1}^{\ast }c_{2}e^{-2it}+\left\vert
c_{2}\right\vert ^{2}\right) ^{1/2}  \label{ZM}
\end{equation}%
and%
\begin{eqnarray}
&&\alpha =i\frac{c_{1}c_{2}^{\ast }e^{2it}-c_{1}^{\ast }c_{2}e^{-2it}}{%
2\left\vert z\right\vert ^{2}},  \label{ZA} \\
&&\beta =\frac{\beta _{0}}{\left\vert z\right\vert }=\pm \frac{\sqrt{%
\left\vert c_{1}\right\vert ^{2}-\left\vert c_{2}\right\vert ^{2}}}{%
\left\vert z\right\vert },  \label{ZB} \\
&&\gamma =\gamma _{0}-\frac{1}{2}\arg z,  \label{ZC} \\
&&\delta =\frac{\beta _{0}}{2\left\vert z\right\vert }\left( c_{3}e^{i\arg
z}+c_{3}^{\ast }e^{-i\arg z}\right) ,  \label{ZD} \\
&&\varepsilon =\frac{i}{2}\left( c_{3}e^{i\arg z}-c_{3}^{\ast }e^{-i\arg
z}\right) ,  \label{ZE} \\
&&\kappa =\kappa _{0}-\frac{i}{8}\left[ c_{3}^{2}\left( 1-e^{2i\arg
z}\right) -\left. c_{3}^{\ast }\right. ^{2}\left( 1-e^{-2i\arg z}\right) %
\right] .  \label{ZF}
\end{eqnarray}%
The inverse relations between the essential, real and complex, parameters
are given by%
\begin{equation}
\alpha _{0}=\frac{i}{2}\left( c_{1}c_{2}^{\ast }-c_{1}^{\ast }c_{2}\right)
,\qquad \beta _{0}=\pm \sqrt{\left\vert c_{1}\right\vert ^{2}-\left\vert
c_{2}\right\vert ^{2}},  \label{ZAB}
\end{equation}%
\begin{equation}
\delta _{0}=\pm \frac{1}{2}\sqrt{\left\vert c_{1}\right\vert ^{2}-\left\vert
c_{2}\right\vert ^{2}}\left( c_{3}+c_{3}^{\ast }\right) ,\qquad \varepsilon
_{0}=\frac{i}{2}\left( c_{3}-c_{3}^{\ast }\right) .  \label{ZDE}
\end{equation}%
These formulas (\ref{ZA})--(\ref{ZF}) provide a complex parametrization of
the Schr\"{o}dinger group for the simple harmonic oscillator originally
found in Ref.~\cite{Niederer73} (see also \cite{Lop:Sus:VegaGroup}, \cite%
{LopSusVegaHarm} and the references therein). A similar parametrization for
the wavefunctions (\ref{WaveFunctionN}) was used in Ref.~\cite{Dod:Man79}
(see \cite{Har:Ben-Ar:Mann11} and \cite{Kr:Sus12} for an extension to
generalized harmonic oscillators).


\begin{thebibliography}{999}
\bibitem{Abadetal11} J.~Abadie et al.,\emph{\ A gravitational wave
observatory operating beyond the quantum shot-noise limit\/}, Nature Physics 
\textbf{7} (2011), 962--965.

\bibitem{AbramVolUFN} E.~G.~Abramochkin and V.~G.~Volostnikov, \emph{Spiral
light beams\/}, Physics--Uspekhi \textbf{47} (2004)~\#~12, 1177--1203.

\bibitem{Agrawaletal74} G.~P.~Agrawal, A.~K.~Ghatak, and C.~L.~Mehtav, \emph{%
Propagation of a partially coherent beam through selfoc fibers\/}, Opt.
Comm. \textbf{12} (1974)~\#~3, 333--337.

\bibitem{Akh:Ber} A.~Akhiezer and V.~B.~Berestetskii, \textsl{Quantum
Electrodynamics\/}, Interscience Publishers, New York, 1965.

\bibitem{AndersonPlus72} R.~L.~Anderson, S. ~Kumei, and C.~E.~Wulfman, \emph{%
Invariants of the equations of wave mechanics. I\/}, Rev. Mex. F\'{\i}s. 
\textbf{21} (1972), 1--33.

\bibitem{AndersonII72} R.~L.~Anderson, S. ~Kumei, and C.~E.~Wulfman, \emph{%
Invariants of the equations of wave mechanics. II One-particle Schr\"{o}%
dinger equations\/}, Rev. Mex. F\'{\i}s. \textbf{21} (1972), 35--57.

\bibitem{Bag:Bel:Ter83} V.~G.~Bagrov, V.~V.~Belov and I.~M.~Ternov, \emph{%
Quasiclassical trajectory-coherent states of a particle in an arbitrary
electromagnetic field\/}, J.~Math. Phys. \textbf{24} (1983)~\#~12,
2855--2859.

\bibitem{Bailey48} W.~N.~Bailey, \emph{Some integrals involving Hermite
polynomials\/}, J.~London Math. Soc. \textbf{23} (1948)~\#~4, 291--297.

\bibitem{Belov:Karavaev1987} V.~V.~Belov and A.~G.~Karavaev, \emph{Higher
approximations for quasiclassical trajectory-coherent states\/}, Izvestiya
Vysshikh Uchebynkh Zavedenij Fizika, \textbf{31} (1987)~\#~10, 14--18 [in
Russian]; see also English transl.: Sov. Phys. Journal 1989, \textbf{30}%
~\#10, 819--822.

\bibitem{Ber:Lif:Pit} V.~B.~Berestetskii, E.~M.~Lifshitz, and
L.~P.~Pitaevskii, \textsl{Relativistic Quantum Theory\/}, Pergamon Press,
Oxford, 1971.

\bibitem{Bia:Bia75} I.~Bia\l ynicki-Birula and Z.~Bia\l ynicki-Birula, 
\textsl{Quantum Electrodynamics\/}, Pergamon Press Ltd. and PWN--Polish
Scientific Publishers, Oxford, New York, Toronto, Sydney, Warszawa, 1975.

\bibitem{BirrelDavies82} N.~D.~Birrell and P.~C.~W.~Davies, \textsl{Quantum
Fields in Curved Space\/}, Cambridge University Press, Cambridge, 1982.

\bibitem{Bouchouleetal99} I.~Bouchoule, H.~Perrin, A.~Kuhn, M.~Morinaga, and
C.~Salomon, \emph{\ Neutral atoms prepared in Fock states of a
one-dimensional harmonic potential\/}, Phys. Rev. A \textbf{59 }(1999)~\#~1,
R8--R11.

\bibitem{Buchleitneretal02} A.~Buchleitner, D.~Delandea, and J.~Zakrzewski, 
\emph{Non-dispersive wave packets in periodically driven quantum systems\/},
Phys. Rep. \textbf{368} (2002), 409--547.

\bibitem{Bo:Shi} N.~N.~Bogoliubov and D.~V.~Shirkov, \textsl{Introduction to
the Theory of Quantized Fields\/}, third edition, John Wiley \& Sons, New
York, Chichester, Brisbane, Toronto, 1980.

\bibitem{BoySharpWint} C.~P. Boyer, R.~T.~Sharp, and P.~Winternitz, \emph{%
Symmetry breaking interactions for the time dependent Schr\"{o}dinger
equation\/}, J.~Math. Phys. \textbf{17} (1976)~\#~8, 1439--1451.

\bibitem{Breit:Schill:Mlyn97} G.~Breitenbach, S.~Schiller, and J.~Mlynek, 
\emph{Measurement of the quantum states of squeezed light\/},~Nature \textbf{%
387} (1997)~May 29, 471--475.

\bibitem{ChernegaManko11} V.~N.~Chernega and V.~I.~Man'ko, \emph{Probability
representation and state-extended uncertainty relations\/}, Journal of
Russian Laser Research \textbf{32} (2011)~\#~2, 125--129.

\bibitem{Cirac12} J.~I.~Cirac, \emph{Entanglement in many-body quantum
systems\/}, arXiv:1205.3742v1 [quant-ph] 16 May 2012.

\bibitem{CiracBlattetal93} J.~I.~Cirac, R.~Blatt, A.~S.~Parkins, and
P.~Zoller, \emph{Preparation of Fock states by observing of quantum jumps in
an ion trap\/}, Phys. Rev. Lett. \textbf{70} (1993)~\#~6, 762--765.

\bibitem{CiracBlattetal94} J.~I.~Cirac, R.~Blatt, A.~S.~Parkins, and
P.~Zoller, \emph{Quantum collapse and revival in the motion of a single
trapped ion\/}, Phys. Rev. A. \textbf{49} (1994)~\#~2, 1202--1207.

\bibitem{CiracBlattZoller94} J.~I.~Cirac, R.~Blatt, and P.~Zoller, \emph{%
Nonclassical states of motion in a three-dimensional ion trap by adiabatic
passage\/}, Phys. Rev. A. \textbf{49} (1994)~\#~5, R3174--R3177.

\bibitem{Ciracetal93} J.~I.~Cirac, A.~S.~Parkins, R.~Blatt, and P.~Zoller, 
\emph{\textquotedblleft Dark\textquotedblright\ states of the motion of a
trapped ion\/}, Phys. Rev. Lett. Lett. \textbf{70} (1993)~\#~5, 556--559.

\bibitem{CiracZoller95} J.~I.~Cirac and P.~Zoller, \emph{Quantum
computations with cold trapped ions\/}, Phys. Rev. Lett. \textbf{74}
(1995)~\#~20, 4091--4094.

\bibitem{Chumakovetal03} S.~M.~Chumakov, A.~B.~Klimov, and M.~Kozierowski, 
\emph{From the Jaynes--Cummings model to collective interactions}, in: 
\textsl{Theory of Nonclassical States of Light\/} (V.~V.~Dodonov and
V.~I.~Man'ko, Editors), Taylor \& Francis, London and New York, 2003.

\bibitem{Cooketal85} R.~J.~Cook, D.~G.~Shankland, and A.~L.~Wells, \emph{%
Quantum theory of particle motion in a rapidly oscillating field\/}, Phys.
Rev. A \textbf{31} (1985)~\#~2, 564--567.

\bibitem{Cor-Sot:Lop:Sua:Sus} R.~Cordero-Soto, R.~M.~L\'{o}pez, E.~Suazo,
and S.~K.~Suslov, \emph{Propagator of a charged particle with a spin in
uniform magnetic and perpendicular electric fields\/}, Lett.~Math.~Phys. 
\textbf{84} (2008)~\#~2--3, 159--178.

\bibitem{Cor-SotSuaSusDamped} R.~Cordero-Soto, E.~Suazo, and S.~K.~Suslov, 
\emph{Models of damped oscillators in quantum mechanics\/}, J. Phys. Math. 
\textbf{1} (2009), S090603 (16 pages).

\bibitem{Cor-Sot:Sua:SusInv} R.~Cordero-Soto, E.~Suazo, and S.~K.~Suslov, 
\emph{Quantum integrals of motion for variable quadratic Hamiltonians\/},
Ann. Phys. \textbf{325} (2010)~\#~9, 1884--1912.

\bibitem{Demkov09} Yu.~N.~Demkov, \emph{Channeling, superfocusing, and
nuclear reactions\/}, Physics of Atomic Nuclei, \textbf{72} (2009)~\#~5,
779--785.

\bibitem{DemkovMeyer04} Yu.~N.~Demkov and J.~D.~Meyer, \emph{A sub-atomic
microscope, superfocusing in channeling and close encounter atomic and
nuclear reactions\/}, Eur. Phys. J. B \textbf{42} (2004), 361--365.

\bibitem{Diedrichetal89} F.~Diedrich, J.~C.~Bergquist, W.~M.~Itano, and
D.~J.~Wineland, \emph{Laser cooling to the zero-point energy of motion\/},
Phys. Rev. Lett. \textbf{62} (1989)~\#~4, 403--406.

\bibitem{DiracQM} P.~A.~M.~Dirac, \textsl{The Principles of Quantum
Mechanics\/}, third edition, Clarendon Press, Oxford, 1947.

\bibitem{Dodonov02} V.~V.~Dodonov, \emph{`Nonclassical' states in quantum
optics: a `squeezed' review of the first 75 years\/}, J. Opt. B: Quantum
Semiclass. Opt. \textbf{4} (2002), R1--R33.

\bibitem{Dodonov10} V.~V.~Dodonov, \emph{Current status of dynamical Casimir
effect\/}, Physica Scripta \textbf{82} (2010)~\#~3, 038105 (10~pp).

\bibitem{DodKlimMan90} V.~V.~Dodonov, A.~B.~Klimov, and\ V.~I.~Man'ko, \emph{%
Generation of squeezed states in a resonator with a moving wall\/}, Phys.
Lett. A. \textbf{149} (1990)~\#~4, 225--228.

\bibitem{Dod:Klim:Nik93} V.~V.~Dodonov, A.~B.~Klimov, and D.~E.~Nikonov, 
\emph{Quantum phenomena in nonstationary media\/}, Phys. Rev. A. \textbf{47}
(1993)~\#~5, 4422--4429.

\bibitem{Dod:Kur:Man80} V.~V.~Dodonov, A.~B.~Kurmyshev, and V.~I.~Man'ko, 
\emph{Generalized uncertainty relation and correlated coherent states\/},
Phys. Lett. A. \textbf{79} (1980)~\#~2,3, 150--152.

\bibitem{Dod:Mal:Man75} V.~V.~Dodonov, I.~A.~Malkin, and V.~I.~Man'ko, \emph{%
Integrals of motion, Green functions, and coherent states of dynamical
systems\/}, Int.~J.~Theor.~Phys. \textbf{14} (1975)~\#~1, 37--54.

\bibitem{Dod:Man79} V.~V.~Dodonov and V.~I.~Man'ko, \emph{Coherent states
and the resonance of a quantum damped oscillator\/}, Phys.~Rev.~A \textbf{20}
(1979)~\#~2, 550--560.

\bibitem{Dodonov:Man'koFIAN87} V.~V.~Dodonov and V.~I.~Man'ko, \emph{%
Invariants and correlated states of nonstationary quantum systems}, in: 
\textsl{Invariants and the Evolution of Nonstationary Quantum Systems\/},
Proceedings of Lebedev Physics Institute, vol. 183, pp. 71-181, Nauka,
Moscow, 1987 [in Russian]; English translation published by Nova Science,
Commack, New York, 1989, pp. 103-261.

\bibitem{DodonovManko03} V.~V.~Dodonov and V.~I.~Man'ko, \emph{%
\textquotedblleft Nonclassical\textquotedblright\ states in quantum optics:
brief historical review}, in: \textsl{Theory of Nonclassical States of
Light\/} (V.~V.~Dodonov and V.~I.~Man'ko, Editors), Taylor \& Francis,
London and New York, 2003.

\bibitem{Dod:Man:Man94} V.~V.~Dodonov, O.~V.~Man'ko, and V.~I.~Man'ko, \emph{%
Photon distribution for one-mode mixed light with a generic gaussian Wigner
function}, Phys. Rev. A \textbf{49} (1994), 2993--3001.

\bibitem{Dunnetal95} T.~J.~Dunn, I.~A.~Walmsley, and S.~Mukamel, \emph{%
Experimental determination of the quantum-mechanical state of a molecular
vibrational mode using fluorescence tomography}, Phys. Rev. Lett. \textbf{74}
(1995), 884--887.

\bibitem{DutraQED} S.~M.~Dutra, \textsl{Cavity Quantum Electrodynamics : The
Strange Theory of Light in a Box\/}, Hoboken, NJ, USA: Wiley, 2005.

\bibitem{Eberleetal10} T.~Eberle et al., \emph{Quantum enhancement of the
zero-area Sagnac interferometer topology for gravitational wave detection\/}%
, Phys. Rev. Lett. \textbf{104} (2010), 251102 (4 pages).

\bibitem{Ehrenfest} P.~Ehrenfest, \emph{Bemerkung \"{u}ber die angen\"{a}%
herte G\"{u}ltigkeit der klassischen Mechanik innerhalb der Quantenmechanik\/%
}, Zeitschrift f\"{u}r Physik~A \textbf{45} (1927), 455--457.

\bibitem{Eichleretal11} C.~Eichler, D.~Bozyigit, C.~Lang, M.~Baur,
L.~Steffen, J.~M.~Fink, S.~Filipp, and A.~Wallraff, \emph{Observation of
two-mode squeezing in the microwave frequency domain}, Phys. Rev. Lett. 
\textbf{107 }(2011)~, 113601 (5 pages).

\bibitem{Ermakov} V.~P.~Ermakov, \emph{Second-order differential equations.
Conditions of complete integrability\/}, Universita Izvestia Kiev, Series
III \textbf{9} (1880), 1--25; see also Appl. Anal. Discrete Math. \textbf{2}
(2008)~\#~2, 123--145 for English translation of Ermakov's original paper.

\bibitem{FeynmanFundamental} R.~P.~Feynman, \textsl{The Theory of
Fundamental Processes\/}, Perseus Books Publishing, Cambridge,
Massachusetts,1998.

\bibitem{FeynmanQED} R.~P.~Feynman, \textsl{QED: The Strange Theory of Light
and Matter\/}, Princeton University Press, Princeton and Oxford, 2006.

\bibitem{Fey:Hib} R.~P.~Feynman and A.~R.~Hibbs, \textsl{Quantum Mechanics
and Path Integrals\/}, McGraw--Hill, New York, 1965.

\bibitem{FermiRad} E.~Fermi, \emph{Quantum theory of radiation\/}, Rev. Mod.
Phys. \textbf{4} (1932), 87--132.

\bibitem{Fermi} E.~Fermi, \textsl{Notes on Quantum Mechanics\/}, Phoenix
Science Series, The University of Chicago Press, Chicago and London, 1961.

\bibitem{Flu} S.~Fl\"{u}gge, \textsl{Practical Quantum Mechanics\/},
Springer--Verlag, Berlin, 1999.

\bibitem{Fock28-2} V.~Fock, \emph{On the relation between the integrals of
the quantum mechanical equations of motion and the Schr\"{o}dinger wave
equation\/}, Zs. Phys. \textbf{49} (1928)~\#~5--6, 323--338; reprinted in:
V.~A.~Fock, \textsl{Selected Works: Quantum Mechanics and Quantum Field
Theory\/}, (L.~D.~Faddeev, L.~A.~Khalfin, and I.~V.~Komarov, Eds.), Chapman
\& Hall/CRC, Boca Raton, London, New York, Washington, D.~C., 2004,
pp.~33--49.

\bibitem{Fock32-2} V.~Fock, \emph{Configuration space and second
quantization\/}, Zs. Phys. \textbf{75} (1932)~\#~9--10, 622--647; reprinted
in: V.~A.~Fock, \textsl{Selected Works: Quantum Mechanics and Quantum Field
Theory\/}, (L.~D.~Faddeev, L.~A.~Khalfin, and I.~V.~Komarov, Eds.), Chapman
\& Hall/CRC, Boca Raton, London, New York, Washington, D.~C., 2004,
pp.~191--220.

\bibitem{Fock34-3} V.~Fock, \emph{On quantum electrodynamics\/}, Phys. Zs.
Sowjetunion \textbf{6} (1934), 425--469; reprinted in: V.~A.~Fock, \textsl{%
Selected Works: Quantum Mechanics and Quantum Field Theory\/},
(L.~D.~Faddeev, L.~A.~Khalfin, and I.~V.~Komarov, Eds.), Chapman \&
Hall/CRC, Boca Raton, London, New York, Washington, D.~C., 2004,
pp.~331--368.

\bibitem{Fu:Mat:Hat:Kur:Zeil} T.~Fujii, Sh.~Matsuo, N.~Hatakenaka,
S.~Kurihara, and A.~Zeilinger, \emph{Quantum circuit analog of the dynamical
Casimir effect\/}, Phys. Rev. B. \textbf{84} (2011)~\#~17, 174521 (9 pages).

\bibitem{Garraway00} B.~M.~Garraway, \emph{Extended Gaussian wavepacket
dynamics\/}, J.~Phys. B: At. Mol. Opt. Phys. \textbf{33} (2000), 4447--4467.

\bibitem{Gerhardtetal09} I.~Gerhardt, G.~Wrigge, G.~Zumofen, J.~Hwang,
A.~Renn, and V.~Sandoghdar, \emph{Coherent state preparation and observation
of Rabi oscillations in a single molecule\/}, Phys. Rev. A. \textbf{79}
(2009)~\#~1, 011402(R) (4 pages).

\bibitem{GlauberCollect} R.~J.~Glauber, \textsl{Quantum Theory of Optical
Coherence: Selected Papers and Lectures\/}, WILEY-VCH Verlag GmbH \& Co.
KGaA, Weinheim, 2007.

\bibitem{Glauber91} R.~J.~Glauber and M.~Lewenstein, \emph{Quantum optics of
dielectric media\/}, Phys. Rev.~A \textbf{43} (1991)~\#~1, 467--491.

\bibitem{Gold:Krivch} I.~I.~Gol'dman and V.~D.~Krivchenkov, \textsl{Problems
in Quantum Mechanics\/}, Dover, New York, 1993.

\bibitem{Guerr:Lop:Ald:Coss11} J.~Guerrero, and F.~F.~L\'{o}pez-Ruiz,
V.~Aldaya, and F.~Cossio, \emph{Harmonic states for the free particle\/}, J.
Phys. A: Math. Theor. \textbf{44} (2011), 445307 (16pp); see also
arXiv:1010.5525v3 [quant-ph] 1 Jul 2011.

\bibitem{Hagen72} C.~H.~Hagen, \emph{Scale and conformal transformations in
Galilean-covariant field theory\/}, Phys. Rev. D \textbf{5} (1972)~\#~2,
377--388.

\bibitem{Har:Ben-Ar:Mann11} G.~Harari, Ya.~Ben-Aryeh, and Ady Mann, \emph{%
Propagator for the general time-dependent harmonic oscillator with
application to an ion trap\/},~Phys. Rev. A \textbf{84} (2011)~\#~6, 062104
(4 pages).

\bibitem{HarocheRaimond06} S.~Haroche J.-M.~Raimond, \textsl{Exploring the
Quantum: Atoms, Cavities, and Photons\/}, Oxford University Press, Oxford,
2006.

\bibitem{HeisenbergQM} W.~Heisenberg, \textsl{The Physical Principles of the
Quantum Theory\/}, University of Chicago Press, Chicago, 1930; Dover, New
York, 1949.

\bibitem{Henry:Glotzer88} R.~W.~Henry and S.~C.~Glotzer, \emph{A
squeezed-state primer\/}, Am. J. Phys. \textbf{56} (1988)~\#~4, 318--328.

\bibitem{Hawking74} S.~W.~Hawking, \emph{Black hole explosions?\/}, Nature,
London \textbf{248} (1974), 30--31.

\bibitem{Hawking75} S.~W.~Hawking, \emph{Particle creation by black holes\/}%
, Commun. Math. Phys. \textbf{43} (1975)~\#~3, 199--220.

\bibitem{HeinzenWineland90} D.~J.~Heinzen and D.~J.~Wineland, \emph{%
Quantum-limited cooling and detection of radio-frequency oscillations by
laser-cooled ions\/}, Phys. Rev. A (1990)~\#~5, 2977--2994.

\bibitem{HilletyetalWigner84} M.~Hillery, R.~F.~O'Connel, M.~O.~Scully, and
E.~P.~Wigner, \emph{Distribution functions in physics: fundamentals\/},
Phys. Rep. \textbf{106} (1953)~\#~3, 121--167.

\bibitem{Hollenhorst79} J.~N.~Hollenhorst, \emph{Quantum limits on
resonant-mass gravitational-radiation detectors\/}, Phys. Rev. D
(1979)~\#~6, 1669--1679.

\bibitem{Husimi53} K.~Husimi, \emph{Miscellanea in elementary quantum
mechanics: I--II\/}, Prog. Theor. Phys. \textbf{9} (1953)~\#~3, 238--244;
Prog. Theor. Phys. \textbf{9} (1953)~\#~4, 381--402.

\bibitem{HusimiOtuka53} K.~Husimi and M.~\^{O}tuka, \emph{Miscellanea in
elementary quantum mechanics: III\/}, Prog. Theor. Phys. \textbf{10}
(1953)~\#~2, 173--190.

\bibitem{It:Zu} C.~Itzykson and J-B.~Zuber, \textsl{Quantum Field Theory\/},
Dover Publications, New York, 2005.

\bibitem{JACKIW80} R.~Jackiw, \emph{Dynamical symmetry of the magnetic
monopole}, Ann. Phys. \textbf{129 }(1980), 183--200.

\bibitem{Jacobson04} T.~A.~Jacobson, \emph{Introduction to quantum fields in
curved spacetime and the Hawking effect\/}, arXiv:0308048v3 [gr-qc] 9 April
2004.

\bibitem{JainetalLv10} N.~Jain, S.~R.~Huisman, E.~Bimbard, and
A.~I.~Lvovsky, \emph{A bridge between the single-photon and squeezed-vacuum
states}, Optics Express \textbf{18 }(2010)~\#~17, 18254 (6 pages).

\bibitem{JaynesCummings63} E.~T.~Jaynes and F.~W.~Cummings, \emph{Comparison
of quantum and semiclassical radiation theories with application to the beam
maser\/,} Proc. IEEE. \textbf{51} (1963)~\#~1, 89--109.

\bibitem{Jessenetal92} P.~S.~Jessen, C.~Gerz, P.~D.~Lett, W.~D.~Phillips,
S.~L.~Rolston, R.~J.~C.~Spreeuw, and C.~I.~Westbrook, \emph{Observation of
quantized motion of Rb atoms in an optical field\/}, Phys. Rev. Lett. 
\textbf{69} (1992)~\#~1, 49--52.

\bibitem{Johanningetal09} M.~Johanning, A.~F.~Var\'{o}n, and Ch.~Wunderlich, 
\emph{Quantum simulations with cold trapped ions\/}, J. Phys. B: At. Mol.
Opt. Phys.\textbf{\ 42} (2009), 154009 (27pp).

\bibitem{KalninsMiller74} E.~G.~Kalnins and W.~Miller, \emph{Lie theory and
separation of variables. 5. The equations }$iU_{t}+U_{xx}=0$ \emph{and }$%
iU_{t}+U_{xx}-c/x^{2}U=0\text{\emph{\/}},$ J. Math. Phys. \textbf{15}
(1974)~\#~10, 1728--1737.

\bibitem{Kennard27} E.~H.~Kennard, \emph{Zur Quantenmechanik einfacher
Bewegungstypen\/}, Zeitschrift f\"{u}r Physik \textbf{44} (1927)~\#~4--5,
326--352.

\bibitem{Kimetal89} M.~S.~Kim, F.~A.~M.~de~Oliveira, and P.~L.~Knight, \emph{%
Properties of squeezed number states and squeezed thermal states}, Phys.
Rev. A \textbf{40} (1989)~\#~5, 2494--2503.

\bibitem{Klauder60} J.~R.~Klauder, \emph{The design of radar signals having
both high range resolution and high velocity resolution\/}, The Bell system
technical journal \textbf{39 }(1960)~\#~4, 809--820.

\bibitem{Klauder12} J.~R.~Klauder, \emph{Enhanced quantization: a primer\/},
J. Phys. A: Math. Theor. \textbf{45 }(2012), 285304 (8 pages).

\bibitem{KlauderSudarshan} J.~R.~Klauder and E.~C.~G.~Sudarshan, \textsl{%
Fundamentals of Quantum Optics\/}, W.~A.~Benjamin, Inc., New York,
Amsterdam, 1968.

\bibitem{Kouchan11} C.~Koutschan, http://hahn.la.asu.edu/\symbol{126}%
suslov/curres/index.htm; see \textsl{Mathematica} notebook: Koutschan.nb;
see also http://www.risc.jku.at/people/ckoutsch/pekeris/

\bibitem{KouchanZeilberger10} C.~Koutschan and D.~Zeilberger. \emph{The 1958
Pekeris-Accad-WEIZAC Ground-Breaking Collaboration that computed Ground
States of Two-Electron Atoms (and its 2010 Redux)\/}, Math. Intelligencer 
\textbf{33} (2011)~\#~2, 52--57.

\bibitem{Kr:Sus12} C. Krattenthaler, S.~I.~Kryuchkov, A.~Mahalov, and
S.~K.~Suslov, \emph{On the problem of electromagnetic-field quantization\/},
arXiv:1301.7328v2 [math-ph] 9 Apr 2013.

\bibitem{KrSusVegaWignerMath} S.~I.~Kryuchkov, S.~K.~Suslov and
J.~M.~Vega-Guzm\'{a}n, http://hahn.la.asu.edu/\symbol{126}%
suslov/curres/index.htm; see \textsl{Mathematica} notebook: WignerSummary.nb.

\bibitem{Lahetal11} P.~L\"{a}hteenm\"{a}ki, G.~S.~Paraoanu, J.~Hassel, and
P.~J.~Hakonen, \emph{Dynamical Casimir effect in a Josephson metamaterial\/}%
, arXiv:1111.5608v2 [cond-mat.mes-hall] 1 Dec 2011.

\bibitem{La:Lif} L.~D.~Landau and E.~M.~Lifshitz, \textsl{Quantum Mechanics:
Nonrelativistic Theory\/}, Pergamon Press, Oxford, 1977.

\bibitem{Lan:Sus} N.~Lanfear and S.~K.~Suslov, \emph{The time-dependent Schr%
\"{o}dinger equation, Riccati equation and Airy functions\/},
arXiv:0903.3608v5 [math-ph] 22 Apr 2009.

\bibitem{Lan:Lop:Sus} N.~Lanfear, R.~M.~L\'{o}pez, and S.~K.~Suslov, \emph{%
Exact wave functions for generalized harmonic oscillators\/}, Journal of
Russian Laser Research \textbf{32} (2011)~\#~4, 352--361; see also
arXiv:11002.5119v2 [math-ph] 20 Jul 2011.

\bibitem{Leach:Andrio08} P. G. L. Leach and K. Andriopoulos, \emph{The
Ermakov equation: a commentary\/}, Appl. Anal. Discrete Math. \textbf{2}
(2008)~\#~2, 146--157.

\bibitem{LeavenSalaMayato01} C.~R.~Leavens, R.~Sala Mayato, \emph{On
constructing the wave function of a quantum particle from its Wigner
phase-space distribution function\/}, Phys. Lett. A \textbf{280} (2001),
163--172.

\bibitem{LeibfriedetalWineland03} D.~Leibfried, R.~Blatt, C.~Monroe, and
D.~Wineland, \emph{Quantum dynamics of single trapped ions\/}, Rev. Mod.
Phys. \textbf{75} (2003) \#~1, 281--324.

\bibitem{LeonardPaul95} U.~Leonhard and H.~Paul, \emph{Measuring the quantum
state of light\/}, Prog. Quant. Electr. \textbf{19} (1995), 89--130.

\bibitem{Lop:Sus} R.~M.~L\'{o}pez and S.~K.~Suslov, \emph{The Cauchy problem
for a forced harmonic oscillator\/}, Revista Mexicana de F\'{\i}sica, 
\textbf{55} (2009)~\#~2, 195--215; see also arXiv:0707.1902v8 [math-ph] 27
Dec 2007.

\bibitem{Lop:Sus:VegaGroup} R.~M.~L\'{o}pez, S.~K.~Suslov, and
J.~M.~Vega-Guzm\'{a}n, \emph{Reconstracting the Schr\"{o}dinger groups\/},
Physica Scripta \textbf{87}\ (2013)~\#~3, 038112 (6 pages); see also \emph{%
On the harmonic oscillator group\/}, arXiv:1111.5569v2 [math-ph] 4 Dec 2011.

\bibitem{LopSusVegaHarm} R.~M.~L\'{o}pez, S.~K.~Suslov, and J.~M.~Vega-Guzm%
\'{a}n, \emph{On a hidden symmetry of quantum harmonic oscillators\/},
Journal of Difference Equations and Applications, \textbf{19} (2013)~\#~4,
543--554:
http://www.tandfonline.com/doi/abs/10.1080/10236198.2012.658384\#preview;
see also arXiv:1112.2586v2 [quant-ph] 2 Jan 2012.

\bibitem{Lop:Sus:VegaMath} R.~M.~L\'{o}pez, S.~K.~Suslov, and J.~M.~Vega-Guzm%
\'{a}n, http://hahn.la.asu.edu/\symbol{126}suslov/curres/index.htm; see 
\textsl{Mathematica} notebook: HarmonicOscillatorGroup.nb.

\bibitem{Lord49} R.~D.~Lord, \emph{Some integrals involving Hermite
polynomials\/}, J.~London Math. Soc. \textbf{24} (1948) \#~2, 101--112.

\bibitem{Louisell73} W.~H.~Louisell, \textsl{Quantum Statistical Properties
of Radiation\/}, Wiley, New York, 1973.

\bibitem{LvovskyBabichev02} A.~I.~Lvovsky and S.~A.~Babichev, \emph{%
Synthesis and tomographic characterization of the displaced Fock state of
light\/}, Phys. Rev. \textbf{66} (2002), 011801(R) (4 pages).

\bibitem{LvovskyHansenetal01} A.~I.~Lvovsky, H.~Hansen, T.~Aichele,
O.~Benson, J.~Mlynek, and S.~Schiller, \emph{Quantum state reconstruction of
the single-photon Fock state\/}, Phys. Rev. Lett. \textbf{87} (2001)~\#~5,
050402 (4 pages).

\bibitem{LvovskyMlynek02} A.~I.~Lvovsky and J.~Mlynek, \emph{Quantum-optical
catalysis: generating nonclassical states of light by means of linear
optics\/}, Phys. Rev. Lett. \textbf{88} (2002)~\#~25, 250401 (4 pages).

\bibitem{LvRay09} A.~I.~Lvovsky and M.~G.~Raymer, \emph{Continuous-variable
optical quantum-state tomography\/}, Rev. Mod. Phys. \textbf{81} (2009),
January--March, 299--332.

\bibitem{MahSusParOptics} A.~Mahalov and S.~K.~Suslov, \emph{Solution of
paraxial wave equation for inhomogeneous media in linear and quadratic
approximation\/}, submitted.

\bibitem{Malkin:Man'ko79} I.~A.~Malkin and V.~I.~Man'ko, \textsl{Dynamical
Symmetries and Coherent States of Quantum System\/}, Nauka, Moscow, 1979 [in
Russian].

\bibitem{Man'koCasimir} V.~I.~Man'ko, \emph{The Casimir effect and quantum
vacuum generator\/}, Journal of Soviet Laser Research \textbf{12} (1991),
383--385.

\bibitem{Marhic78} M.~E.~Marhic, \emph{Oscillating Hermite--Gaussian wave
functions of the harmonic oscillator\/}, Lett. Nuovo Cim. \textbf{22}
(1978)~\#~8, 376--378.

\bibitem{Meekhofetal96} D.~M.~Meekhof, C.~Monroe, B.~E.~King, W.~M.~Itano,
and D.~J.~Wineland, \emph{Generation of nonclassical motional states of a
trapped atom\/}, Phys. Rev. Lett. \textbf{76 }(1996)~\#~11, 1796--1799.

\bibitem{Monroeetal95} C.~Monroe, D.~M.~Meekhof, B.~E.~King, W.~M.~Itano,
and D.~J.~Wineland, \emph{Demonstration of a fundamental quantum logic gate\/%
}, Phys. Rev. Lett. \textbf{75 }(1995)~\#~25, 4714--4717.

\bibitem{Merz} E.~Merzbacher, \textsl{Quantum Mechanics\/}, third edition,
John Wiley \& Sons, New York, 1998.

\bibitem{Miller77} W.~Miller, Jr., \textsl{Symmetry and Separation of
Variables\/}, Encyclopedia of Mathematics and Its Applications, Vol.~4,
Addison--Wesley Publishing Company, Reading etc, 1977.

\bibitem{Morinagaetal99} M.~Morinaga, I.~Bouchoule, J.-C.~Karam, and
C.~Salomon, \emph{Manipulation of motional quantum states of neutral atoms\/}%
, Phys. Rev. Lett. \textbf{83} (1999)~\#~20, 4037--4040.

\bibitem{Moyal47} J.~E.~Moyal, \emph{Quantum mechanics as a statistical
theory\/}, Proc. Camb. Phil. Soc. \textbf{49} (1947), 99--124.

\bibitem{Nationetal12} P.~D.~Nation, J.~R.~Johansson, M.~P.~Blencowe, and
F.~Nori, \emph{Stimulationg uncertainty: Amplifying the quantum vacuum with
superconducting circuits\/}, Rev. Mod. Phys. \textbf{84} (2012),
January--March, 1--24.

\bibitem{Niederer72} U.~Niederer, \emph{\ The maximal kinematical invariance
group of the free Schr\"{o}dinger equations\/}, Helv. Phys. Acta \textbf{45}
(1972), 802--810.

\bibitem{Niederer73} U.~Niederer, \emph{The maximal kinematical invariance
group of the harmonic oscillator\/}, Helv. Phys. Acta \textbf{46} (1973),
191--200.

\bibitem{Ni:Su:Uv} A.~F.~Nikiforov, S.~K.~Suslov, and V.~B.~Uvarov, \textsl{%
Classical Orthogonal Polynomials of a Discrete Variable\/},
Springer--Verlag, Berlin, New York, 1991.

\bibitem{Ourjoumtsevetal11} A.~Ourjoumtsev, A.~Kubanek, M.~Koch, C.~Sames,
P.~W.~H.~Pinkse, G.~Rempe, and K.~Murr, \emph{Observation of squeezed light
from one atom excited with two photons\/}, Nature \textbf{474 }(2011),
623--626.

\bibitem{Paul90} W.\ Paul, \emph{\ Electromagnetic traps for charged and
neutral particles\/}, Rev. Mod. Phys. \textbf{62 }(1990)~\#~3, 531--540.

\bibitem{PerelomovCSBook} A.~M.~Perelomov, \textsl{Generalized Coherent
States and Their Applications\/}, Springer--Verlag, Berlin, 1986.

\bibitem{Pikovskietal12} I.~Pikovski, M.~R.~Vanner, M.~Aspelmeyer,
M.~S.~Kim, and \v{C}.~Brukner, \emph{Probing Planck-scale physics with
quantum optics\/}, Nature Physics \textbf{8 }(2012), 393--397.

\bibitem{RiesetalLv03} J.~Ries, B.~Brezger, and A.~I.~Lvovsky, \emph{%
Experimental vacuum squeezing in rubidium vapor via self-rotation\/}, Phys.
Rev. A \textbf{68 }(2003), 025801 (4 pages).

\bibitem{Rempeetal90} G.~Rempe, F.~Schmidt-Kaler, and H.~Walther, \emph{%
Observation of sub-Poissonian photon statistics in a micromaser\/},
Phys.~Rev. Lett. \textbf{64} (1990)~\#~23, 2783--2786.

\bibitem{RempeWaltherKlein87} G.~Rempe, H.~Walther, and N.~Klein, \emph{%
Observation of quantum collapse and revival in a one-atom maser\/},
Phys.~Rev. Lett. \textbf{58} (1987)~\#~4, 353--356.

\bibitem{Rosen76} S.~Rosencrans, \emph{Perturbation algebra of an elliptic
operator\/}, J.~Math. Anal. Appl. \textbf{56} (1976)~\#~2, 317--329.

\bibitem{Roosetal99} Ch.~Roos, Th.~Zeiger, H.~Rohde, H.~C. N\"{a}gerl,
J.~Eschner, D.~Leibfried, F.~Schmidt-Kaler, and R.~Blatt, \emph{Quantum
state engineering on an optical transition and decoherence in a Paul trap\/}%
, Phys.~Rev. Lett. \textbf{83} (1999)~\#~23, 4713--4716.

\bibitem{Rowe77} D.~J.~Rowe, \emph{The two-photon laser beam as a breathing
mode of the electromagnetic field\/}, Can. J. Phys. \textbf{56 }(1978),
442--446.

\bibitem{SanSusVin} B.~Sanborn, S.~K.~Suslov, and L.~Vinet, \emph{Dynamic
invariants and Berry's phase for generalized driven harmonic oscillators\/},
Journal of Russian Laser Research \textbf{32} (2011)~\#~5, 486--494; see
also arXiv:1108.5144v1 [math-ph] 25 Aug 2011.

\bibitem{Santosetal07} M.~F.~Santos, G.~Giedke, and E.~Solano, \emph{%
Noise-free measurement of marmonic oscillators with instantaneous
interactions\/}, Phys. Rev. Lett. \textbf{98 }(2007), 020401 (4 pages).

\bibitem{Schiff} L.~I.~Schiff, \textsl{Quantum Mechanics\/}, third edition,
McGraw-Hill, New York, 1968.

\bibitem{SchillerBreitenbachetal96} S.~Schiller, G.~Breitenbach,
S.~F.~Pereira, T.~M\"{u}ller, and J.~Mlynek, \emph{Quantum statistics of the
squeezed vacuum by measurement of the density matrix in the number state
representation\/}, Phys. Rev. Lett. \textbf{77 }(1996)~\#~14, 2933--2936.

\bibitem{Schleich01} W.~P.~Schleich, \textsl{Quantum Optics in Phase Space\/}%
, Wiley--Vch Publishing Company, Berlin etc, 2001.

\bibitem{Schradeetal95} G.~Schrade, V.~I.~Man'ko, W.~P.~Schleich, and
R.~J.~Glauber, \emph{Wigner functions in the Paul trap\/}, Quantum
Semiclass. Opt. B \textbf{7} (1995)~\#~3, 307--325.

\bibitem{SchroedingerOscillator} E.~ Schr\"{o}dinger, \emph{Quantisierung
als Eigenwertproblem II\/}, Annalen der Physik, \textbf{79} (1926), 489-527;
see also \textsl{Collected Papers on Wave Mechnics\/}, Blackie \& Son Ltd,
London and Glascow, 1928, pp.~13--40, for English translation of Schr\"{o}%
dinger's original paper.

\bibitem{Schroedinger} E.~ Schr\"{o}dinger, \emph{Der stetige \"{U}bergang
von der Mikro-zur Makro Mechanik\/}, Die Naturwissenshaften, \textbf{28}
(1926), 664--666; see also \textsl{Collected Papers on Wave Mechnics\/},
Blackie \& Son Ltd, London and Glascow, 1928, pp.~41--44, for English
translation of Schr\"{o}dinger's original paper.

\bibitem{Sch:Plun:Soff98} R.~Sch\"{u}tzhold, G.~Plunien, and G.~Soff, \emph{%
Trembling cavities in the canonical approach\/}, Phys. Rev. A \textbf{57 }%
(1998)~\#~4, 2311--2318.

\bibitem{Shchukinetal09} E.~Shchukin, Th.~Kiesel, and W.~Vogel,\emph{\
Generalized minimum-uncertainty squeezed states\/}, Phys. Rev. A \textbf{79 }%
(2009), 043831 (7 pages).

\bibitem{ShoreKnight93} B.~W.~Shore and P.~L.~Knight, \emph{The
Jaynes--Cummings model}, J. Mod. Opt. \textbf{40} (1993)~\#~7, 1195--1238.

\bibitem{Slater1950} J.~C.~Slater, \textsl{Microvawe Electronics\/}, D. van
Nostrand Co. Inc., New York, chapter 4, 1950.

\bibitem{Slusheretal85} R.~E.~Slusher, L.~W.~Hollberg, B. Yurke,
J.~C.~Mertz, and J.~F.~Valleys, \emph{Observation of squeezed states
generated by four-wave mixing in an optical cavity}, Phys. Rev. Lett. 
\textbf{55 }(1985)~\#~22, 2409--2412.

\bibitem{Stenholm86} S.~Stenholm, \emph{Amplification of squeezed states\/},
Optics Communications \textbf{58 }(1986)~\#~3, 177--180.

\bibitem{Stoler70} D.~Stoler, \emph{Equivalence Classes of Minimum
Uncertainty Packets\/}, Phys. Rev. D \textbf{1 }(1970)~\#~12, 3217--3219.

\bibitem{Stoler71} D.~Stoler, \emph{Equivalence Classes of Minimum
Uncertainty Packets. II\/}, Phys. Rev. D \textbf{4 }(1971)~\#~6, 1925--1926.

\bibitem{SuazoSuslovSol} E.~Suazo and S.~K.~Suslov, \emph{Soliton-like
solutions for nonlinear Schr\"{o}dinger equation with variable quadratic
Hamiltonians\/}, Journal of Russian Laser Research \textbf{33} (2012)~\#~1,
63--82; arXiv:1010.2504v4 [math-ph] 24 Nov 2010.

\bibitem{SuazoSusVega10} E.~Suazo, S.~K.~Suslov, and J.~M.~Vega-Guzm\'{a}n, 
\emph{The Riccati equation and a diffusion-type equation\/}, New York J.
Math. \textbf{17a} (2011), 225--244.

\bibitem{SuazoSusVega11} E.~Suazo, S.~K.~Suslov, and J.~M.~Vega-Guzm\'{a}n, 
\emph{The Riccati system and a diffusion-type equation\/}, arXiv:
1102.4630v1 [math-ph] 22 Feb 2011.

\bibitem{Suslov10} S.~K.~Suslov, \emph{Dynamical invariants for variable
quadratic Hamiltonians\/}, Physica Scripta \textbf{81} (2010)~\#~5, 055006
(11~pp); see also arXiv:1002.0144v6 [math-ph] 11 Mar 2010.

\bibitem{Suslov11} S.~K.~Suslov, \emph{On integrability of nonautonomous
nonlinear Schr\"{o}dinger equations\/}, Proc. Amer. Math. Soc. \textbf{140}
(2012)~\#~9, 3067--3082; see also arXiv:1012.3661v3 [math-ph] 16 Apr 2011.

\bibitem{Suslov12} S.~K.~Suslov, \emph{An analog of the Berry phase for
simple harmonic oscillators\/}, Physica Scripta \textbf{87} (2013)~\#~3,
038118 (4 pages); arXiv:1112.2418v1 [quant-ph] 12 Dec 2011.

\bibitem{SuslovMath} S.~K.~Suslov, http://hahn.la.asu.edu/\symbol{126}%
suslov/curres/index.htm; see \textsl{Mathematica} notebook: BerrySummary.nb.

\bibitem{Takabayasi54} T. Takabayasi, \emph{The formulation of quantum
mechanics in terms of ensemble in phase space\/}, Prog. Theor. Phys. \textbf{%
11} (1954)~\#~4, 341--373.

\bibitem{Unruh76} W.~G.~Unruh, \emph{Notes on black-hole evaporation\/},
Phys. Rev. D \textbf{14} (1976)~\#~4, 870--892.

\bibitem{Vahlbruch08} H.~Vahlbruch, M.~Mehmet, S.~Chelkowski et al, \emph{%
Observation of squeezed light with 10-dB quantum-noise reduction\/}, Phys.
Rev. Lett. \textbf{100} (2008), 033602 (4 pages).

\bibitem{Verkerketal92} P.~Verkerk, B.~Lounis, C.~Salomon, and
C.~Cohen-Tannoudji, \emph{Dynamics and spatial order of cold Cesium atoms in
a periodic optical potential\/}, Phys. Rev. Lett. \textbf{68} (1992)~\#~26,
3861--3864.

\bibitem{VinRudSuxBook79} M.~B.~Vinogradova, O.~V.~Rudenko, and
A.~P.~Sukhorukov, \textsl{Theory of Waves\/}, Nauka, Moscow, 1979 [in
Russian].

\bibitem{VogelFilho95} W.~Vogel and R.~L.~de Matos Filho, \emph{Nonlinear
Jaynes-Cummings dynamics of a trapped ion\/}, Phys. Rev. A \textbf{52}
(1995)~\#~5, 4214--4217.

\bibitem{Vysotskii2Adamenko12} V.~I.~Vysotskii, M.~V.~Vysotskyy, and
S.~V.~Adamenko, \emph{Formation and application of correlated states in
nonstationary systems at low energies of interacting particles\/}, J. Exp.
Theor. Phys.--JETP \textbf{114} (2012)~\#~2, 243--252.

\bibitem{Walls83} D.~F.~Walls, \emph{Squeezed states of light\/},~Nature 
\textbf{306} (1983)~November~10, 141--146.

\bibitem{WallsMilburn08} D.~F.~Walls and G.~J.~Milburn, \textsl{Quantum
Optics\/}, Springer, Berlin, Heidelberg, 2008.

\bibitem{Wein} S.~Weinberg, \textsl{The Quantum Theory of Fields\/}, volumes
1--3, Cambridge University Press, Cambridge, 1998.

\bibitem{WeinbergQM} S.~Weinberg, \textsl{Lectures on Quantum Mechanics\/},
Cambridge University Press, New York, 2013.

\bibitem{Wigner32} E. Wigner, \emph{On the quantum correction for
thermodynamic equilibrium\/}, Phys. Rev. \textbf{40} (1932), 749--759.

\bibitem{Wilsonetal11} C.~M.~Wilson, G.~Johansson, A.~Pourkabirian,
M.~Simoen, J.~R.~Johansson, T.~Duty, F.~Nori, and P.~Delsing, \emph{%
Observation of the dynamical Casimir effect in a superconducting circuit\/}%
,~Nature \textbf{479} (2011)~November~17, 376--379.

\bibitem{You:Nori11} J.~Q.~You and F.~Nori, \emph{Atomic physics and quantum
optics using superconducting circuits\/}, Nature \textbf{474} (2011),
589--597.

\bibitem{Yuen76} H.~P.~Yuen, \emph{Two-photon coherent states of the
radiation field\/}, Phys. Rev. A \textbf{13} (1976)~\#~6, 2226--2243.
\end{thebibliography}
\end{document}